\documentclass[aps,showpacs,floats,prb,amsmath,amssymb,twocolumn]{revtex4}
\usepackage{graphicx}
\usepackage{bm}
\usepackage[applemac]{inputenc}

\graphicspath{{pdffig/}{epsfig/}}

\def\be{\begin{equation}}
\def\ee{\end{equation}}

\def\bi{\begin{itemize}}
\def\ei{\end{itemize}}
\def\bn{\begin{enumerate}}
\def\en{\end{enumerate}}
\def\bea{\begin{eqnarray}}
\def\eea{\end{eqnarray}}
\def\no{\nonumber}
\def\ba{\begin{array}}
\def\ea{\end{array}}
\def\bd{\begin{displaymath}}
\def\ed{\end{displaymath}}

\def\la{\langle}
\def\ra{\rangle}

\def\pa{\parallel}
\def\pe{\perp}

\newcommand{\bS}{{\bf S}}
\newcommand{\bh}{{\bf h}}
\newcommand{\bk}{{\bf k}}
\newcommand{\bz}{{\bf z}}
\newcommand{\bq}{{\bf q}}
\newcommand{\bQ}{{\bf Q}}

\newcommand{\hh}{\hat{h}}

\newcommand{\LiX}{Li$_{\text 2}$VO$X$O$_{\text 4}$}
\newcommand{\AAVO}{$AA'$VO(PO$_{\text 4}$)$_{\text 2}$}

\begin{document}

\title{Magnetocaloric effect in the frustrated square lattice
J$_{\text 1}$-J$_{\text 2}$ model}

\author{B. Schmidt and P. Thalmeier} \affiliation{Max Planck Institute
for the Chemical Physics of Solids, 01187 Dresden, Germany}

\author{Nic Shannon} \affiliation{H.H.\ Wills Physics Laboratory,
University of Bristol, Tyndall Avenue, Bristol BS8 1TL, UK}

\begin{abstract}
    We investigate the magnetocaloric properties of the
    two-dimensional frustrated $J_1$-$J_2$ model on a square lattice.
    This model describes well the magnetic behavior of two classes of
    quasi-two-dimensional $S=1/2$ vanadates, namely the \LiX{}
    ($X=\text{Si},\text{Ge}$) and \AAVO{}
    ($A,A'=\text{Pb},\text{Zn},\text{Sr},\text{Ba}$) compounds.  The
    magnetocaloric effect (MCE) consists in the adiabatic temperature
    change upon changing the external magnetic field.  In frustrated
    systems, the MCE can be enhanced close to the saturation field
    because of massive degeneracies among low lying excitations.  We
    discuss results for the MCE in the two distinct antiferromagnetic
    regimes of the phase diagram.  Numerical finite temperature
    Lanczos as well as analytical methods based on the spin wave
    expansion are employed and results are compared.  We give explicit
    values for the saturation fields of the vanadium compounds.  We
    predict that at subcritical fields there is first a (positive)
    maximum followed by sign change of the MCE, characteristic of all
    magnetically ordered phases.
\end{abstract}


\maketitle

\section{Introduction}

Two dimensional (2D) magnetic systems are favorite models to study the
influence of quantum fluctuations on magnetic order.  Depending on the
model they may both prohibit an ordered ground state or select a
specific order among classically degenerate states.  These phenomena
have been studied in great detail for geometrically frustrated systems
like trigonal, Kagom\'e and checkerboard lattice~\cite{misguich:04}.
However they are also present in magnets where the frustration is not
the result of lattice geometry but of competition between different
(for example nearest- and next-nearest neighbor) magnetic bonds.  A
prime example is the frustrated $J_1$-$J_2$ model on a square lattice.
Its ground state and thermodynamic properties in zero field have been
well studied (see Refs.~\onlinecite{misguich:04,shannon:04,shannon:06}
and references cited therein).

Classically the model predicts three magnetic phases depending on the
frustration ratio $J_2/J_1$: The ferromagnet (FM), $(\pi,\pi)$ N\'eel
antiferromagnet (NAF) and $(\pi,0)$ collinear antiferromagnet (CAF).
However it is known that close to the classical CAF/NAF and CAF/FM
boundary quantum fluctuations destroy magnetic order and presumably
stabilize nonmagnetic order parameters.

The discovery of two classes of layered vanadium oxides \LiX{}
($X=\text{Si},\text{Ge}$)~\cite{millet:98, melzi:00, melzi:01} and
\AAVO{}
($A,A'=\text{Pb},\text{Zn},\text{Sr},\text{Ba}$)~\cite{kaul:05,kini:06}
which are well described by this model has further raised interest in
the $J_1$-$J_2$ model.  One advantage of the new vanadium compounds is
a comparatively low energy scale for the exchange constants of order
$10\,\text K$.  Therefore high field experiments might be a promising
way to learn more about their physical properties, indeed the
saturation field for these compounds where the fully polarized state
is achieved seems within experimental reach.
 
Therefore in this work we study exhaustively the high-field magnetic
and especially the magnetocaloric effects (MCE) in the   
$J_1$-$J_2$ model. We use a variety of analytical and numerical
techniques to investigate the dependence of magnetization,
susceptibility, entropy specific heat and adiabatic cooling rate on
magnetic field, temperature and frustration ratio. The variation of
the saturation field with the frustration ratio is calculated and
predictions for the abovementioned compounds are made.  We show that
the low temperature specific heat is strongly enhanced around the
classical phase boundaries where large quantum fluctuations occur.

Our special focus is on the magnetocaloric effect.  We will show that
the cooling rate, normalized to its paramagnetic value, is strongly
enhanced above the saturation field and depends on the frustration
angle.  We also predict that for subcritical fields the cooling rate
is first positive with a maximum at moderate fields and a
subsequently changes sign at a larger subcritical field. This
behavior is common to all AF phases of the model and 
can be understood quantitatively from calculations of contours of
constant entropy in the $(h,T)$ plane.  
The dependence of corresponding characteristic fields on the 
frustration ratio are also calculated.

In Sec.~\ref{sect:MCE} we give a brief description of the
magnetocaloric effect in magnets.  In Sec.~\ref{sect:J1J2MODEL} the
basic properties and phase diagram of the $J_1$-$J_2$ model are
introduced.  In Sec.~\ref{sect:LANCZOS} we discuss extensively results
of the finite temperature Lanczos method (FTLM) for finite 2D
$J_1$-$J_2$ clusters.  In Sec.~\ref{sect:ANALYTIC} we use analytical
methods within mean field or spin wave approximation as an alternative
way to study the magnetocaloric properties.  In
Sec.\ref{sect:DISCUSSION} we discuss and compare the results obtained
by the various methods.  Finally Sec.~\ref{sect:CONC} gives the
summary and conclusion.

\section{The Magnetocaloric effect in magnetically ordered compounds}

\label{sect:MCE}

When a crystal containing magnetic ions is placed in a magnetic field
the adiabatic or isentropic change of this external parameter causes a
temperature change in the sample.  This is called the magnetocaloric
effect (MCE) which was first discovered by
Warburg~\cite{warburg:1881}.  It is nowadays interesting in several
aspects.  Firstly suitable compounds, like paramagnetic salts where
demagnetization leads to cooling may be used
technically~\cite{pecharsky:99,tishin:99}.  Secondly at high (pulsed)
fields the magnetocaloric anomalies at a magnetic phase transition may
be used to map out the $H$-$T$ phase diagrams which are not accessible
otherwise.  Finally it has recently gained special attention in
frustrated magnets.  There the behavior around the saturation field
may be described by the condensation of a macroscopic number of local
magnons~\cite{zhitomirsky:03,zhitomirsky:04,zhitomirsky:05} which
leads to a giant enhancement of the magnetocaloric cooling rate.  The
latter is defined as the rate of change of temperature $T$ with
magnetic field $H$ at fixed entropy $S$.  Using a Maxwell relation one
can write this as
\begin{eqnarray}
    \label{COOLRATE}
    \Gamma_{\text{mc}}\equiv
    \left(\frac{\partial T}{\partial H}\right)_S &=& 
    -\frac{\left(\frac{\partial S}{\partial  H}\right)_T}
    {\left(\frac{\partial S}{\partial T}\right)_H}
    = - \frac{T}{C_V}\left( \frac{\partial M}{\partial T}\right)_H,
\end{eqnarray}
where $C_{V}$ is the heat capacity and $M$ the magnetization of the
sample.  The integrated adiabatic temperature change along an
isentropic line with $S(T,H) = \text{const}$ which is caused by the
variation of magnetic field is then given by
\begin{eqnarray}
    \Delta
    T_{\text{ad}}(H_0,H)=T-T_0=
    \int_{H_0}^H\Gamma_{\text{mc}}(H',T'){\rm d}H'
\end{eqnarray}
Here $T_0$ and $H_0$ are the starting values of temperature and field
respectively.  We take the adiabatic cooling rate for free
paramagnetic ions as a reference quantity.  As shown later it is
simply given by $\Gamma^0_{\text{mc}}=(T/H)$.  The dimensionless
magnetocaloric enhancement factor due to interaction effects is then
defined by the ratio
\begin{eqnarray}
    \hat{\Gamma}_{\text{mc}}\equiv\Gamma_{\text{mc}}/
    \Gamma^0_{\text{mc}}=(H/T)\Gamma_{\text{mc}}.
\end{eqnarray}
In the present work we study the MCE on a Heisenberg square lattice
$J_1$-$J_2$ model which incorporates a frustration of nearest- and
next-nearest neighbor exchange interactions.  It is a suitable model
to analyze the magnetic properties of two classes of
quasi-two-dimensional vanadates, namely the \LiX{}
($X=\text{Si},\text{Ge}$)~\cite{millet:98, melzi:00, melzi:01} and
\AAVO{} ($A,A'=\text{Pb},\text{Zn},\text{Sr},\text{Ba}$)
compounds~\cite{kaul:05,kini:06}.  We investigate several aspects of
magnetocaloric properties.  We show that indeed it may be used to
identify the saturation field and the associated $H_{\text c}(T)$
phase boundary between the fully polarized and the AF ordered states.
We also discuss whether it may be used as a diagnostic for the
appropriate frustration angle (or $J_2/J_1$ ratio).  Finally we will
study whether the frustration effect lead to a visible signature in
the anomalies of the magnetocaloric cooling rates, especially close to
the saturation fields.  We will employ both numerical FTLM methods as
well as approximate analytical methods based on mean field and spin
wave approximations.

\section{The $J_1$-$J_2$ Heisenberg model and examples}

\label{sect:J1J2MODEL}

We first give a brief characterization of the  2D square lattice
$J_1$-$J_2$ model in an external field which is defined by the
Hamiltonian
\begin{eqnarray}
\label{HAMJ1J2}
{\cal H} &=& J_1\sum_{\langle ij \rangle_1} {\bf S}_i.{\bf S}_j 
 + J_2\sum_{\langle ij \rangle_2} {\bf S}_i.{\bf S}_j
 - h \sum_i S^z_i
\end{eqnarray}
with the convention that each bond is counted only once.  Here $J_1$
is the nearest-neighbor exchange coupling along the edges and $J_2$
the next-nearest neighbor exchange coupling along the diagonals of
each square.  Furthermore $h=g\mu_{\text B}H$ where $H$ is the applied
magnetic field.  Here $g$ is the gyromagnetic ratio and $\mu_{\text
B}$ the Bohr magneton.  The zero-field phase diagram may best be
characterized by introducing the equivalent parameter set
\begin{eqnarray}
    J_{\text c}=(J^2_1+J^2_2)^\frac{1}{2};\quad \phi=\tan^{-1}(J_2/J_1)
\end{eqnarray}

\begin{figure}
    \vspace{5mm}
    \centering
    \includegraphics[width=0.35\textwidth]{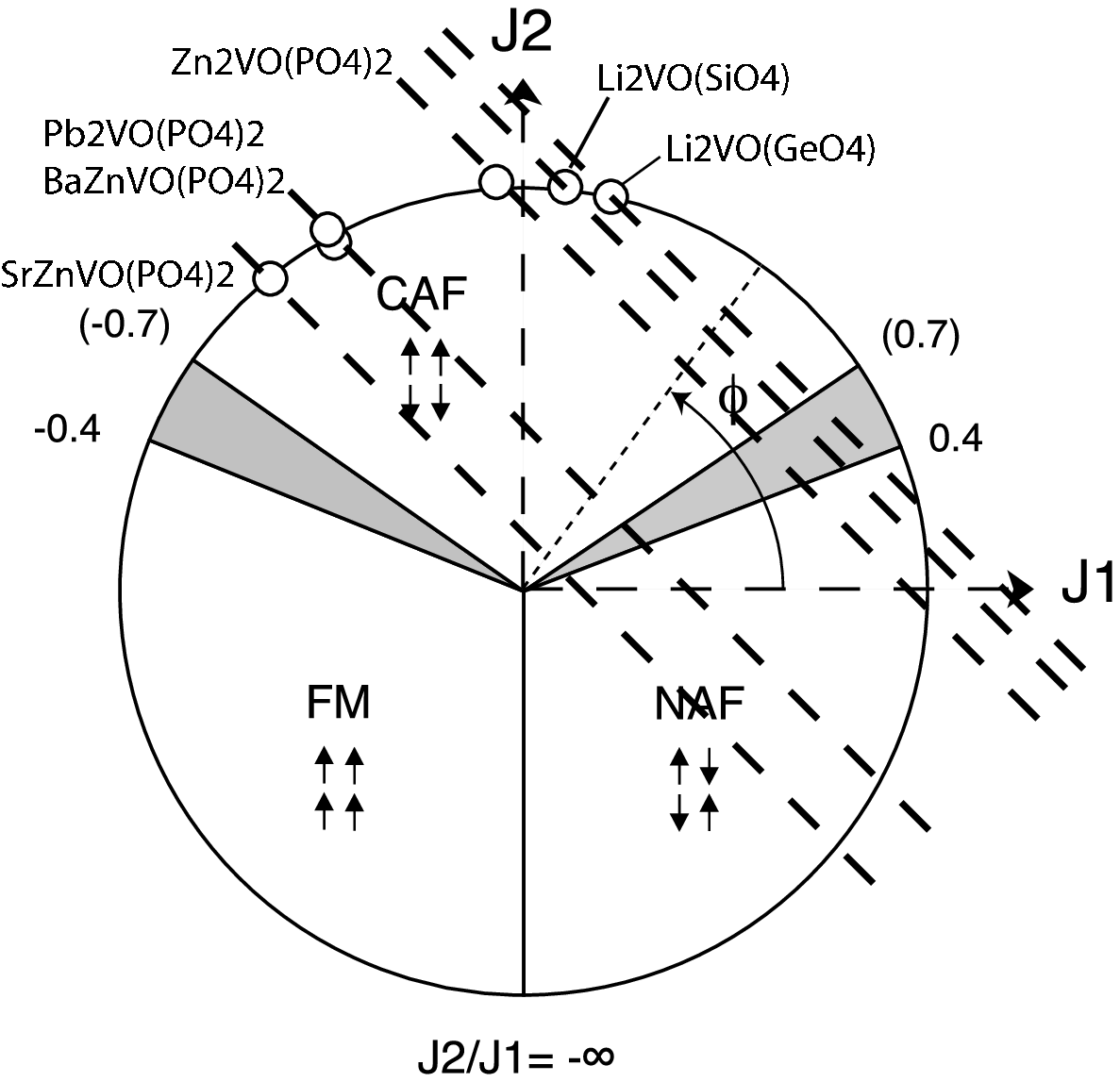}
    \caption{Phases of the spin-1/2 2D square lattice $J_1$-$J_2$
    model.  The CAF and NAF order is indicated by arrows, the
    associated wave vectors are {\bf Q} = (0,1) or (1,0) and (1,1) (in
    units of $\pi$/a) respectively.  The boundary between FM and NAF
    phase is the line $J_{1}=0$, $J_{2}<0$.  Values of $J_{2}/J_{1}$
    in parentheses indicate where zero point fluctuations destroy the
    CAF order parameter~\protect\cite{shannon:06}.  Dashed lines
    correspond to the experimental
    $\Theta_{\text{CW}}=(J_{1}+J_{2})/k_{\text B}$ and refer to the
    known $J_{1}$-$J_{2}$ compounds~\protect\cite{kaul:05,kini:06}.
    Two solutions $\phi_+$ (CAF) and $\phi_-$ (NAF) are compatible
    with the thermodynamic properties.  Here we choose the former
    since they are confirmed by neutron diffraction for the Li- and
    Pb- compounds}
\label{fig:OSphase}
\end{figure}

The `frustration angle' $\phi$ is a convenient quantity to
characterize the amount of exchange frustration in the model and
$J_{\text c}$ gives the energy scale at which thermodynamic anomalies
in specific heat susceptibility etc.  are to be expected.  For
spin-1/2, as function of $\phi$ three main phases (FM,NAF,CAF) appear
already on the classical level
(Fig.~\ref{fig:OSphase})~\cite{shannon:04}.  However for $\phi\sim
0.15\pi$ ($J_2/J_1\sim0.5$) and $\phi\sim 0.85\pi$ ($J_2/J_1\sim-0.5$)
where CAF meets NAF and FM respectively a large degeneracy of the
classical ground state appears and quantum fluctuations lead to
non-magnetic ``hidden order'' phases shown as the shaded sectors in
Fig.~\ref{fig:OSphase}.  These phases have been extensively discussed
in~\cite{misguich:04,shannon:04,shannon:06} and references cited
therein.

For the compounds mentioned above generally \protect\mbox{$J_{\text
c}\simeq10\,\text K$}.  In addition from the high temperature expansion
of the susceptibility the Curie-Weiss temperature is obtained as
$\Theta_{\text CW}=(J_1+J_2)/k_{\text B}$.  Both quantities can be
obtained from experiment and the appropriate pair of exchange
constants ($J_1$, $J_2$) then has to lie on the intersect of a circle
($J_{\text c}$) and a straight line ($\Theta_{\text CW}$) as shown in
Fig.~\ref{fig:OSphase}.  Obviously there are always two solutions
lying in the NAF ($\phi_-$) and CAF ($\phi_+$) sector.  This
observation is unchanged by a more detailed analysis of susceptibility
and specific heat~\cite{misguich:03,shannon:04}.

Various other methods have been proposed to resolve the ambiguity of
frustration angles such as measurement of the spin structure
factor~\cite{shannon:04}, nonlinear susceptibility~\cite{schmidt:05}
and the saturation field of the magnetization~\cite{schmidt:06}.  Only
the former has been tried sofar for Li$_{\text 2}$VO(SiO$_{\text 4}$)
and Pb$_{\text 2}$VO(PO$_{\text 4}$)$_{\text 2}$~\cite{skoulatos:07}.
In both cases the ground state clearly has CAF order.  For this reason
we have assigned the other known family members of $J_1$-$J_2$
vanadates to the same sector in Fig.~\ref{fig:OSphase} although a
confirmation for this conjecture is still lacking.  We believe that
high field investigations are a further promising method to shed light
on these compounds , especially because the comparatively low
$J_{\text c}$ ($\sim 10\,\text K$) will lead to saturation fields
relatively easy to access.

Therefore in this work we study the $J_1$-$J_2$ model in an external
field as given by Eq.~(\ref{HAMJ1J2}).  Thereby we focus on the theory
of the saturation field and the magnetocaloric anomalies both around
the saturation field and for smaller fields within the ordered phase.
We will use both numerical analysis of finite clusters based on the
FTLM method as well as analytical methods based on mean field or
linear spin wave approximations for comparison.  In the latter we
focus on the three main phases with magnetic order.  The analytical
treatment of the magnetocaloric effect in the hidden order phases
warrants a separate treatment which takes into account the proper
non-magnetic order parameter.

\section{Exact diagonalization for finite clusters at finite
temperatures and finite field}
\label{sect:LANCZOS}

We have performed numerical exact-diagonalization calculations for three
different clusters; squares with 16 and 20 sites and a 24-site
rectangle.   All of these tile the lattice in such a way as to be 
compatible with both the $(\pi,\pi)$ NAF and$(\pi,0)$ CAF states, once periodic
boundary conditions are imposed.  Our main focus is on the
finite-temperature, finite-field properties of the $J_{1}$-$J_{2}$ model.  
We therefore use the finite-temperature Lanczos
method (FTLM) to evaluate the partition function of the model, 
together with thermodynamic averages of the form
\begin{eqnarray}
    \left\langle A(T,H)\right\rangle & = & \frac{1}{\cal Z}
    \mathop{\rm Tr}\left(Ae^{-{\cal H}/(k_{\text B}T)}
    \right),
    \label{eqn:trace}  \\
    {\cal Z} & = & \mathop{\rm Tr}e^{-{\cal H}/(k_{\text B}T)},
    \label{eqn:z}
\end{eqnarray}
Here $A$ is an operator, $\cal H$ is the Hamiltonian of the {\em
J}$_{\text 1}$-{\em J}$_{\text 2}$ model including the Zeeman term
(Eq.~(\ref{HAMJ1J2})) and ${\cal Z}(T,H)$ its (field-dependent)
partition function.  For each frustration angle $\phi$, we perform
between 100 and 500 Lanczos iterations with different starting vectors
in each symmetry sector of the Hilbert space.  We keep between 1 and
100 eigenvalues and eigenvectors of the tridiagonal Lanczos matrix per
iteration in order to evaluate the thermodynamic traces discussed in
the following.  Details of the method can be found in
Ref.~\onlinecite{jaklic:00}.

\subsection{Level crossings, spinwave instabilities and saturation 
fields}
\label{sec:levels}

\begin{figure*}
    \centering
    \hfill
    \includegraphics[height=.3\textwidth]{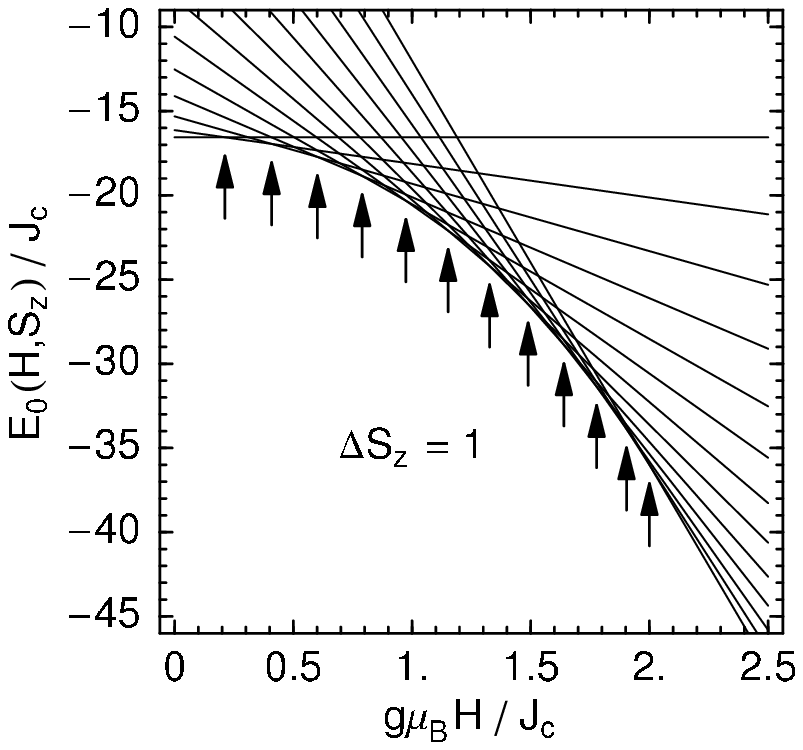}
    \hfill
    \includegraphics[height=.3\textwidth]{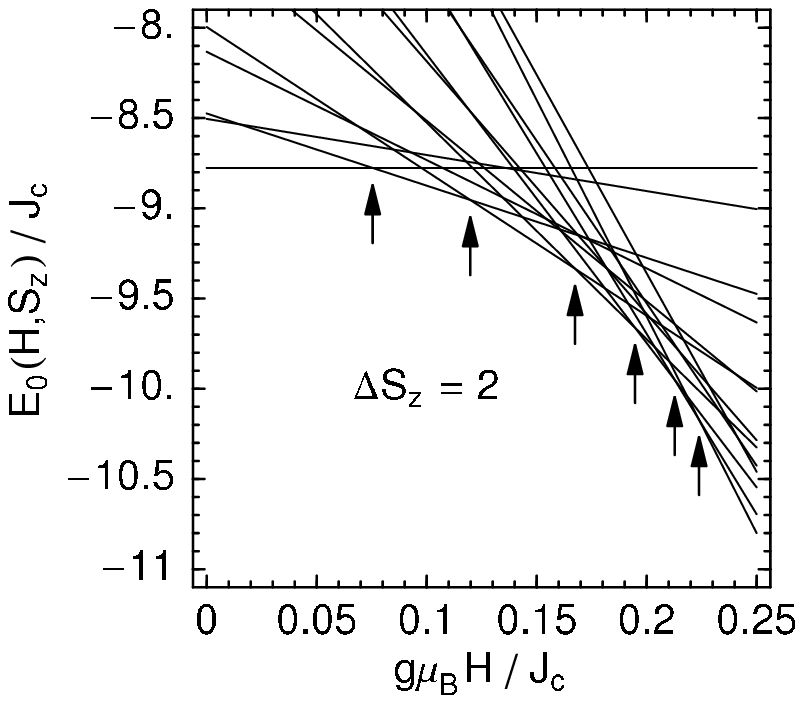}
    \hfill\null \caption{Energy levels as a function of the applied
    magnetic field for two different frustration angles $\phi=0$ (Néel
    antiferromagnet, $J_1 > 0$, left) and $\phi/\pi=0.84$ (collinear
    antiferromagnet, $J_1 < 0$, right).  For each sector of the Hilbert space
    with total $S_{z}=\text{const.}$, the field dependence of the
    respective ground state is plotted.  The arrows point to the
    energy/field values where a jump in the ground-state magnetization
    of the full system occurs.}
    \label{fig:levelcrossings}
\end{figure*}
Before addressing the finite-temperature results, let us examine
certain general features of the model at zero temperature.  Applying a
magnetic field $H$ leads to a Zeeman splitting of the energy levels,
and therefore level crossings occur when increasing the field.  These
level crossings correspond to jumps in the magnetization at zero
temperature, until the fully polarized state is reached at a certain
critical value of the magnetic field.  Figure~\ref{fig:levelcrossings}
illustrates this behavior for two different values of the frustration
angle $\phi$:

As an example for positive (antiferromagnetic) $J_{1}$, we show the
field dependence of the energy levels for the pure Néel
antiferromagnet ($J_{2}=0$) on the left side of the figure.  In the
whole ``right half'' of the phase diagram ($J_{1}>0$, $J_{2}$
arbitrary), the field dependence of the energy levels is qualitatively
similar.  For the 24-site cluster considered, 12 spin flips with
$\Delta S_{z}=1$, indicated by the small arrows, occur at the points
where the magnetic field is given by
\begin{equation}
    g\mu_{\text B}H_{\text{flip}}(S_{z}) = \frac{1}{N}\left(
    E_{0}(S_{z})-E_{0}(S_{z}-1)\right).
    \label{eqn:hflip}
\end{equation}
Here, $E_{0}(S_{z})$ denotes the ground-state energy for the subspace
with constant $S_{z}$ at zero field where $N$ is the cluster size.
The saturation field $H_{\text{sat}}$ is reached when $S_{z}=N/2$.
Because the fully polarized state is an eigenstate of the Hamiltonian,
the numerical values for $H_{\text{sat}}$ from the equation above are
exactly identical to what one finds within linear-spinwave theory for
the infinite system,
\begin{eqnarray}
    \frac{g\mu_{\text B}}{J_{\text c}}H_{\text{sat}}^{\text{LSW}}&=&
    zS\left[\cos\phi\left(1-\frac{1}{2}\left(\cos Q_{x}+\cos 
    Q_{y}\right)\right)\right.
    \nonumber\\
    &&{}
    +\left.\sin\phi\left(\vphantom{\frac{1}{2}}1-\cos Q_{x}\cos
    Q_{y}\right)\right]
    \label{eqn:hsatcl}
\end{eqnarray}
with $z=4$, $S=1/2$, and ${\bf Q}=(\pi,\pi)$ or one of $(\pi,0)$,
$(0,\pi)$ is the antiferromagnetic ordering vector.

In contrast, for ferromagnetic $J_{1}<0$ (but $\phi$ outside the
ferromagnetic 
regime in the phase diagram), we see a qualitatively different 
behavior of the energy levels. The right-hand-side of 
Fig.~\ref{fig:levelcrossings} shows an example for the frustration 
angle $\phi/\pi=0.84$. Instead of 12 level crossings, there are only
six, 
each corresponding to a jump $\Delta S_{z}=2$ occurring at fields
\begin{equation}
    g\mu_{\text B}H_{\text{flip}}^{(k)} = \frac{1}{Nk}\left(
    E_{0}\left(S_{z}\right)
    -E_{0}\left(S_{z}-k\right)\right),\quad k=2.
\end{equation}
The saturation field is now given by an instability criterion of the
fully 
polarized state towards a two-magnon excitation,
\begin{equation}
    g\mu_{\text B}H_{\text{c}k}=\frac{1}{Nk}\left(
    E_{0}\left(\frac{N}{2}\right)
    -E_{0}\left(\frac{N}{2}-k\right)\right),\quad k=2.
    \label{eqn:hsat}
\end{equation}
For $J_{1}<0$, this field is larger than the field of the one-magnon
instability
given by the above equation with $k=1$ and therefore determines the
predominant instability when lowering the field in the fully polarized
state.

A necessary condition for a $\Delta S_{z}=1$ level crossing to occur
is that the lower bound $E_{0}(S_{z})$ of the energy spectrum at zero
field for a fixed value $S_{z}$ is a convex function of $S_{z}$,
i.\,e., the condition
\begin{equation}
    E_{0}(S_{z}+1)\le\frac{1}{2}\left(E_{0}(S_{z})+E_{0}(S_{z}+2)\right)
    \label{eqn:convexity}
\end{equation}
must be fulfilled at $H=0$.  At the special point $J_{1}=0$, $J_{2}>0$
($\phi=\pi/4$), the $J_{1}$-$J_{2}$ lattice decouples into two
independent Néel sublattices, and equality holds above.  For the {\em
finite size\/} clusters which we consider, enlarging $\phi$ further
(i.\,e.{} making $J_{1}$ ferromagnetic) stabilizes two-magnon bound
states, and Eq.~(\ref{eqn:convexity}) no longer holds.  Level
crossings are characterized by $\Delta S^z=2$, and the saturation
field $H_{\text c2}$ is given by Eq.~(\ref{eqn:hsat}) with $k=2$.
This is exactly what would be expected where a spin nematic state is
selected by quantum fluctuations in applied magnetic
field~\cite{shannon:06}.

\begin{figure*}
    \centering
    \hfill
    \includegraphics[height=.32\textwidth]{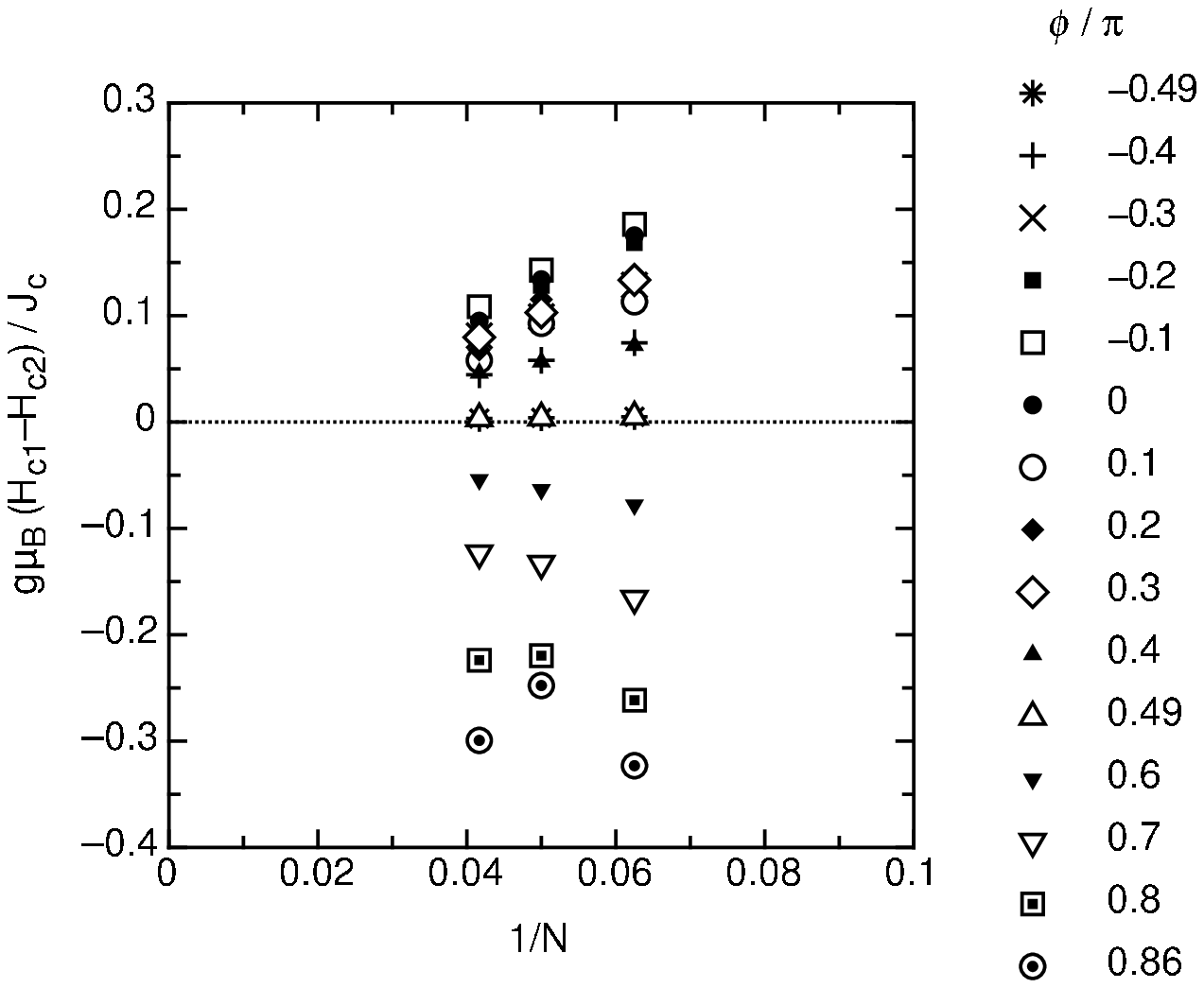}
    \hfill
    \includegraphics[height=.3\textwidth]{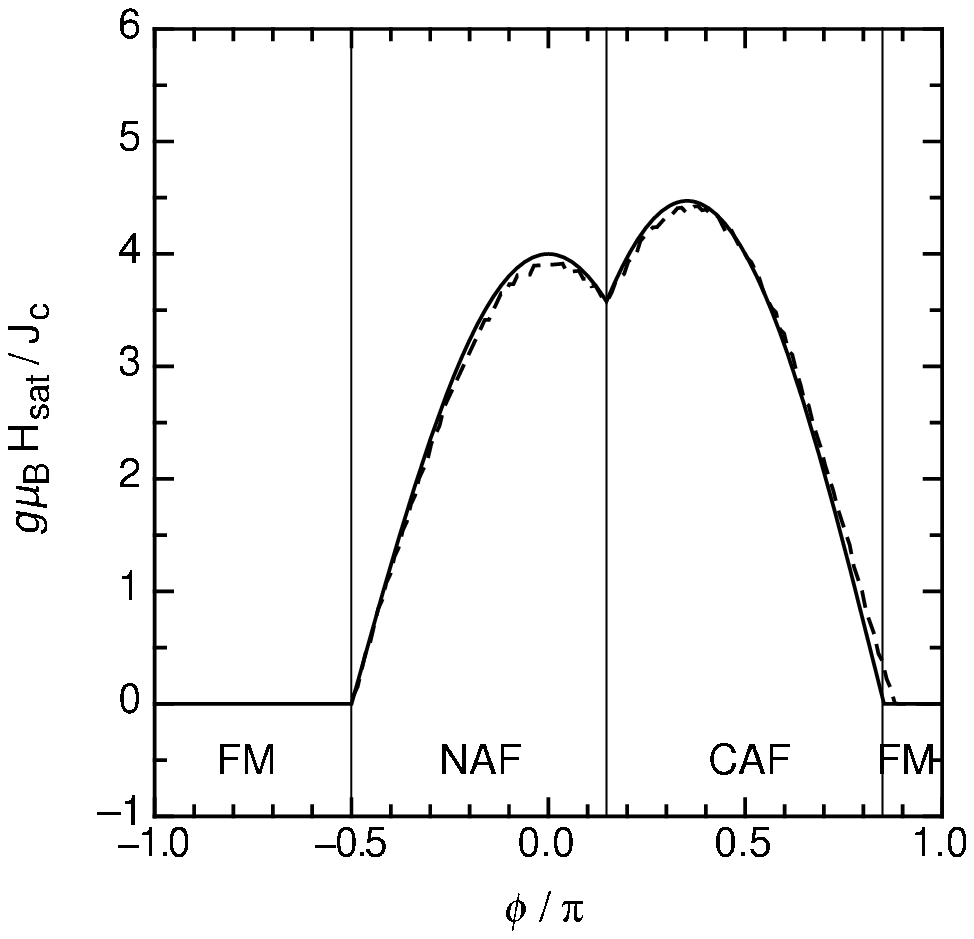}
    \hfill\null \caption{Left: Scaling plot of the difference between
    the one- and two-magnon instability 
    fields, as defined by
    Eq.~(\protect\ref{eqn:hsat}), for cluster sizes $N=16,20$ and
    $24$ sites.  The different symbols denote different positions in
    the phase diagram, corresponding to the values of $\phi$ listed on
    the right-hand side.  Note that the size of the field difference
    is non-monotonous in $\phi$.  Right: One-magnon (solid line) 
    and two-magnon (dashed-line) instability fields for the
    24-site cluster as a function of the frustration angle.  In this
    and in subsequent plots of $\phi$-dependent quantities, the thin
    vertical lines denote the classical phase boundaries of the 
    $J_{1}$-$J_{2}$ model~\protect\cite{shannon:04}.}
    \label{fig:scaling}
\end{figure*}
However these are finite size results, and must be approached with a
little caution.  The critical fields associated with one- and
two-magnon excitations show quite different finite size scaling
properties as a function of $\phi$, as illustrated in
Fig.~\ref{fig:scaling}.  From the three different cluster sizes
studied here, the following observations can be made : Firstly,
$\Delta H_{\text c}$ is a non-monotonic function of the frustration
angle; it has a minimum $\Delta H_{\text c}=0$ at the crossover
between the NAF and CAF phases for $J_{1}=2J_{2}$
($\phi/\pi\approx0.15$).  Secondly, $\Delta H_{\text c}$ changes sign
at $J_{1}=0$ in the CAF phase ($\phi/\pi=1/2$) in favor of a $\Delta
S_{z}=2$ instability as above.  Thirdly, for
$-1/2\le\phi/\pi\lesssim0.8$, $\left|\Delta H_{\text c}\right|$ is a
monotonically decreasing function of $1/N$ and seems to extrapolate to
zero for $N\to\infty$.  This means that the one-magnon instability
(and conventional canted AF order) is restored in the thermodynamic
limit for FM $J_1$ and all $J_2 \gtrsim 0.6|J_1|$.  Only close to the
classical CAF/FM boundary at $J_2 = 0.5|J_1|$ does a two-magnon
instability (with associated nematic order) prevail.  These results
are in complete agreement with previous exact analytic calculations
for two-magnon bound states in the thermodynamic limit, and numerical
exact diagonalizations of larger clusters~\cite{shannon:06}.

\begin{figure}
    \centering
    \includegraphics[width=.35\textwidth]{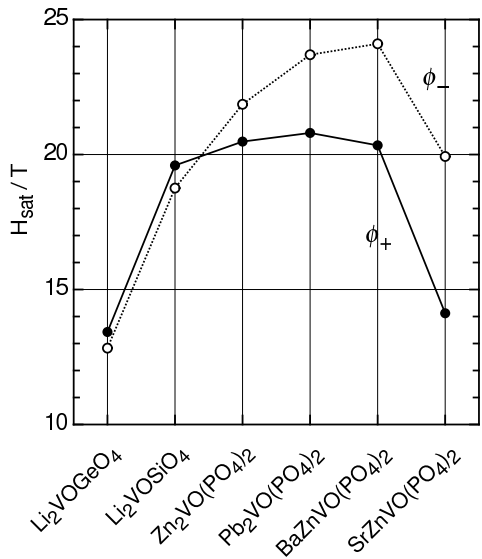}
    \caption{Predicted values for the saturation fields of the
    experimentally known compounds.  $\phi_{+}$ labels the frustration
    angle corresponding to the collinear phase, and $\phi_{-}$ denotes
    the frustration angle for the Néel phase.  The values for
    $\phi_{\pm}$ are determined from zero-field susceptibility and
    heat capacity measurements~\protect\cite{kaul:04,kaul:05,kini:06}.}
    \label{fig:expfields}
\end{figure}
Using the two values $\phi=\phi_{\pm}$ for the frustration angle
together with the experimental energy scale $J_{\text c}$ determined
from zero-field susceptibility and heat capacity
measurements~\cite{shannon:04,kaul:04,kaul:05,kini:06}, we can extract
the expected values for the two different saturation fields
$H_{\text{sat}}=H_{\pm}$ from the right-hand side plot of
Fig.~\ref{fig:scaling}.  Assuming a value $g=2$ for the average
gyromagnetic ratio, we arrive at field values between $13$ and
$24\,\text T$, low enough to be reached experimentally.  In
Fig.~\ref{fig:expfields} we have plotted the predicted values for
$H_{\text{sat}}$ as a function of the frustration angles $\phi_{\pm}$
for the known compounds (Fig.~\ref{fig:OSphase}), using the
corresponding values of $J_{\text c}$ from zero-field measurements.
$H_{\text{sat}}$ can be determined for example by a magnetization
measurement at sufficiently large fields.  For the PO$_{\text
4}$-based compounds, the saturation field together with the zero-field
data for the susceptibility and the heat capacity would provide a
direct way to determine the exchange constants $J_{1}$ and $J_{2}$
individually and hence the region of the phase diagram to which the
compound belongs, without the need to measure the magnetic ordering
vector directly.

\subsection{Magnetization and susceptibility}

We have calculated the magnetization $m(T,H)$ and the magnetic
susceptibility $\chi(T,H)=N_{\text A}\mu_{0}(\partial m(T,H)/\partial
H)$ by evaluating the following thermodynamic traces:
\begin{widetext}
    \begin{eqnarray}
	m(T,H) & = &
	\frac{1}{N}\;g\mu_{\text B}\left\langle S_z^{\text{tot}}\right\rangle,
	\label{eqn:mu}  \\
	\frac{J_{\text c}}{N_{\text A}\mu_0g^2\mu_{\text
	B}^2}\;\chi(T,H)&=&
	\frac{1}{N}\,\frac{J_{\text c}}{k_{\text B}T}
	\left(\left\langle\left(S_z^{\text{tot}}\right)^2\right\rangle -
	\left\langle S_z^{\text{tot}}\right\rangle^2\right),
	\label{eqn:chi}
    \end{eqnarray}
\end{widetext}
where we have explicitly included a factor $1/N$ to account for the
volume dependence of these extensive quantities.  In the definition of
the susceptibility, we also include the magnetic permeability
$\mu_{0}$ and the Avogadro number $N_{\text A}$.  In order to make
$\chi(T,H)$ a dimensionless quantity, we need to multiply it with the
characteristic energy scale $J_{\text c}$.

\begin{figure*}
    \centering
    \hfill
    \includegraphics[height=.35\textwidth]{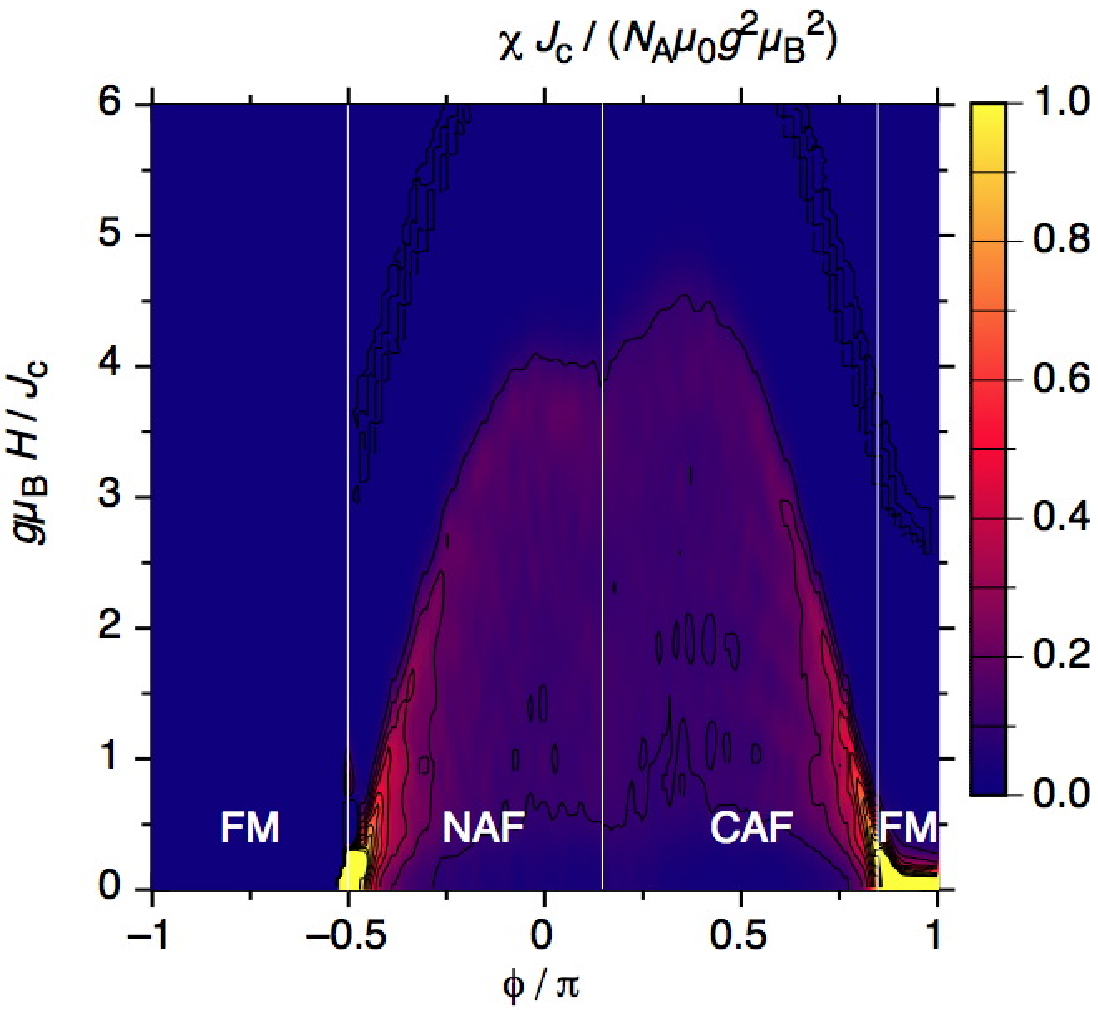}
    \hfill
    \includegraphics[height=.35\textwidth]{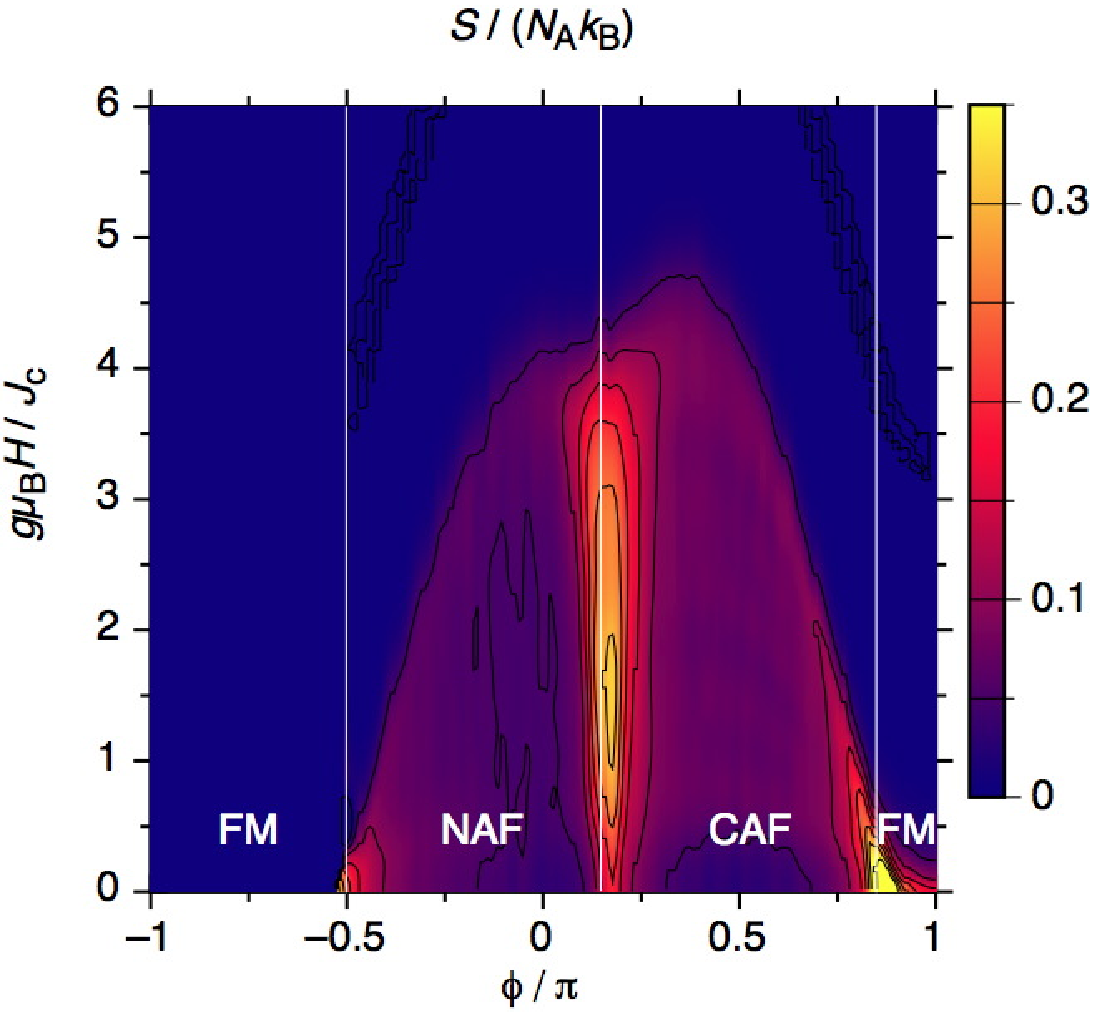}
    \hfill\null \caption{(Color) Contour plots of the magnetic
    susceptibility $\chi(T,H)$ (left) and the entropy $S(T,H)$ (right)
    at a fixed temperature $T=0.2\,J_{\text c}/k_{\text B}$ as a
    function of the frustration angle $\phi$ and the magnetic field
    $H$.  The plot was made using a 24-site cluster on a grid of
    $200\times300$ data points.}
    \label{fig:chiands}
\end{figure*}
At $T=0$, the magnetization $m(T=0,H)$ of any finite size system
evolves as a series of discrete steps.  For a generic AF system with a
singlet ground state, $m(T=0,H)$ takes on all possible (integer) spin
values as a function of $H$, up to the field $H_{\text c}$ at which
the system saturates.  Generally, in the thermodynamic limit,
$m(T=0,H<H_{\text c})$ is a smooth curve, and singular features occur
only where there is a magnetic phase transition.  However in the case
of the spin-1/2 $J_1$-$J_2$ model, a ``step'' at exactly half the
saturation magnetization $m(T=0,H) = 1/2$ survives in the
thermodynamic limit for $J_2 \approx J_1/2$, i.\,e.{} where a
nonmagnetic ground state separates NAF and CAF order.  This
half-magnetization ``plateau'' is believed to be associated with the
formation of localized magnon excitations~\cite{honecker:01}.

Temperature acts to smear jumps in magnetization.  For the small
clusters which we consider, the step-like behavior in $m(T,H)$ has
already disappeared for $k_{\text B}T=0.2\,J_{\text c}$.  At the same
time, all trace os the half-magnetization plateau is also lost.  The
magnetic susceptibility therefore shows a smooth and nearly constant
field dependence, see Fig.~\ref{fig:chiands}.  It drops to zero upon
reaching the saturation field.  Only at the borders of the
ferromagnetic regime for $h=0$ do anomalies related to spontaneous
magnetization appear.

\subsection{Entropy and heat capacity}

The entropy and heat capacity are defined through
\begin{widetext}
    \begin{eqnarray}
	\frac{1}{N_{\text A}k_{\text B}}\;S(T,H) &=&
	\frac{1}{N}\left(\ln {\cal Z}(T,H) +
	\frac{1}{k_{\text B}T}\left\langle{\cal H}(H)\right\rangle\right),
	\label{eqn:entropy}  \\
	\frac{1}{N_{\text A}k_{\text B}}\;C_{V}(T,H) &=& 
	\frac{1}{N}\,\frac{1}{(k_{\text B}T)^2}
	\left(\left\langle{\cal H}^2(H)\right\rangle -
	\left\langle{\cal H}(H)\right\rangle^2
	\right),
	\label{eqn:cv}
    \end{eqnarray}
\end{widetext}
again using the definition for the thermal averages given in
Eqs.~(\ref{eqn:trace},\ref{eqn:z}).  $\cal Z$ and $\cal H$ are the
partition function and the Hamiltonian of Eq.~(\ref{HAMJ1J2}),
respectively.  The right-hand side of Fig.~\ref{fig:chiands} shows a
contour plot of the entropy $S(T,H)$ at fixed temperature $k_{\text
B}T=0.2\,J_{\text c}$ as a function of the frustration angle $\phi$
and the magnetic field $H$.  In the ordered phases, the entropy is a
smooth and almost constant function of the magnetic field, dropping to
$0$ for fields higher than the saturation field $H_{\text{sat}}$.
Characteristic anomalies can be observed at both edges of the
collinear phase, where the entropy crosses a broad maximum as a
function of field before again vanishing when crossing the saturation
field.

\begin{figure*}
    \centering
    \hfill
    \includegraphics[height=.35\textwidth]{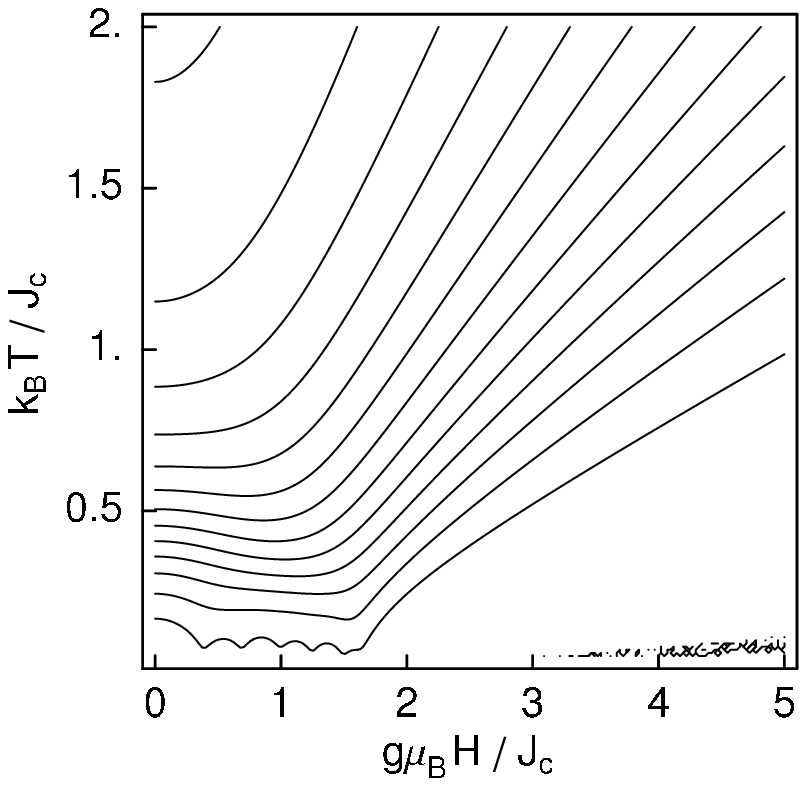}
    \hfill
    \includegraphics[height=.35\textwidth]{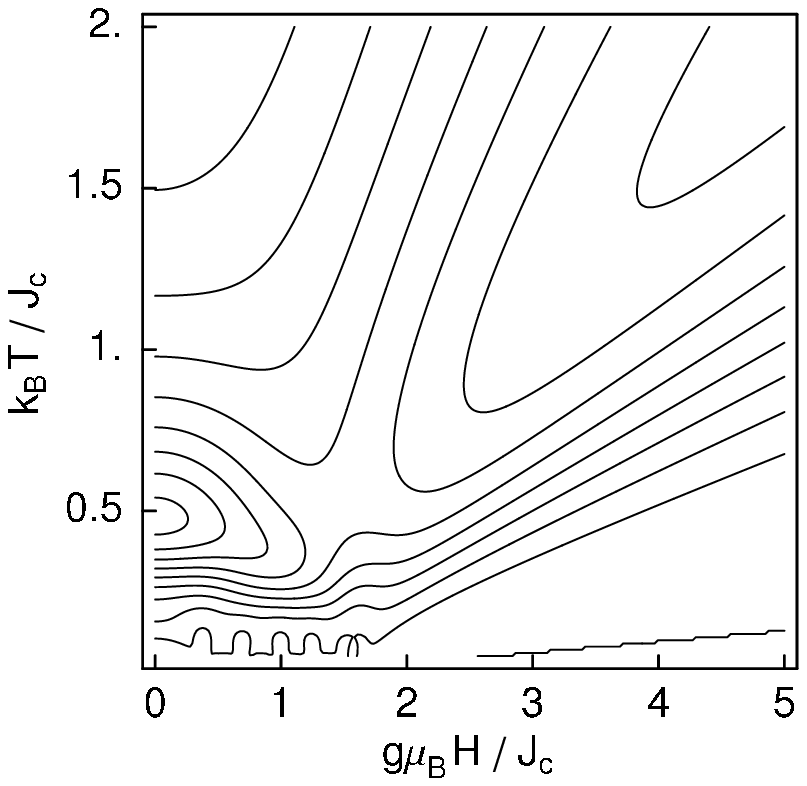}
    \hfill\null \caption{Contour plots of the entropy $S(T,H)$ (left)
    and the heat capacity $C_{V}(T,H)$ (right) at a fixed frustration
    angle $\phi=0.74\,\pi$ as a function of the magnetic field $H$ and
    temperature $T$ for a cluster of 24 sites.  For the entropy plot,
    the contour line starting at $H=0$ and $k_{\text
    B}T\approx0.15\,J_{\text c}$ corresponds to $S=0.05\,N_{\text
    A}k_{\text B}$, the highest contour line starting at $H=0$ and
    $k_{\text B}T\approx1.15\,J_{\text c}$ corresponds to
    $S=0.6\,N_{\text A}k_{\text B}$.  In the plot of the heat capacity
    on the right-hand side, the lowest contour starting at $H=0$ and
    $k_{\text B}\approx0.1\,J_{\text c}$ has a value of
    $C_{V}=0.05\,N_{\text A}k_{\text B}$, while the highest contour
    starts at $H=0$, $k_{\text B}T\approx0.4\,J_{\text c}$ and has
    $C_{V}=0.45\,N_{\text A}k_{\text B}$.}
    \label{fig:isosandcv}
\end{figure*}
Figure~\ref{fig:isosandcv} shows contour plots of the entropy (left)
and heat capacity (right) as a function of magnetic field and
temperature for a fixed value of $\phi=0.747\,\pi$.  We have chosen
this particular frustration angle for the plots because it is believed
to belong to $\phi=\phi_{+}$ for the compound SrZnVO(PO$_{\text
4}$)$_{2}$.  The wiggly contour lines at low temperatures $k_{\text
B}T\ll J_{\text c}$ are finite-size effects, where each temperature
minimum corresponds to a Zeeman level crossing at $T=0$ as discussed
in Section~\ref{sec:levels}.

The lines of constant entropy (Fig.~\ref{fig:isosandcv}, left) are
almost field-independent or even have a slightly negative slope as a
function of field for low temperatures $T\ll J_{\text c}/k_{\text B}$
and fields $H\ll H_{\text{sat}}$.  This implies that a sample {\em
cools down\/} slightly when increasing $H$.  The behavior of a
paramagnet is opposite: Here, isentropic lines are straight lines
crossing the origin, and a sample always {\em heats up\/} when
increasing the applied field.  Of course, for high enough fields and
temperatures, the behavior of the entropy of the $J_{1}$-$J_{2}$
model is the same as that of a paramagnet.  At the saturation field,
which is $H_{\text{sat}}\approx1.64\,J_{\text c}/(g\mu_{\text B})$ for
$\phi/\pi=0.747$, the temperature reaches a minimum when adiabatically
changing the field at low temperatures, and rises steeply when
increasing the field to higher values $H>H_{\text{sat}}$.  In this
area of the phase diagram, a $J_{1}$-$J_{2}$ compound is a good system
for magnetic cooling, especially in view of the low values for the
saturation fields (in Tesla) for the experimentally known compounds
(see Fig.~\ref{fig:expfields} and its discussion above).

The heat capacity $C_{V}(T,H)$ is characterized by two maxima as
shown in the
right panel of Fig.~\ref{fig:isosandcv}.  One maximum occurs at
$H=0$ and $k_{\text B}T\approx0.5\,J_{\text c}$, which is the
broad anomaly occurring at the crossover to the
$1/T^{2}$-temperature dependence for high temperatures: We have
\begin{equation}
    \frac{1}{N_{\text A}k_{\text B}}\;C_{V}(T,H=0)\to
    \left(\frac{J_{\text c}}{k_{\text B}T}\right)^{2},
    \quad T\to\infty.
    \label{eqn:cvhight}
\end{equation}
This maximum has already been discussed in
Ref.~\onlinecite{shannon:04}.  A second maximum can be observed at
high temperatures $T\gg J_{\text c}/k_{\text B}$ and magnetic fields
$H\gg H_{\text {sat}}$.  Here, a field-induced gap opens, leading to a
Schottky-type anomaly of an effective two-level system, see also
Sec.~\ref{sec:MF} and Fig.~\ref{fig:MFTHERMH} (center).

\subsection{The magnetocaloric effect}

\begin{figure*}
    \centering
    \hfill
    \includegraphics[height=.35\textwidth]{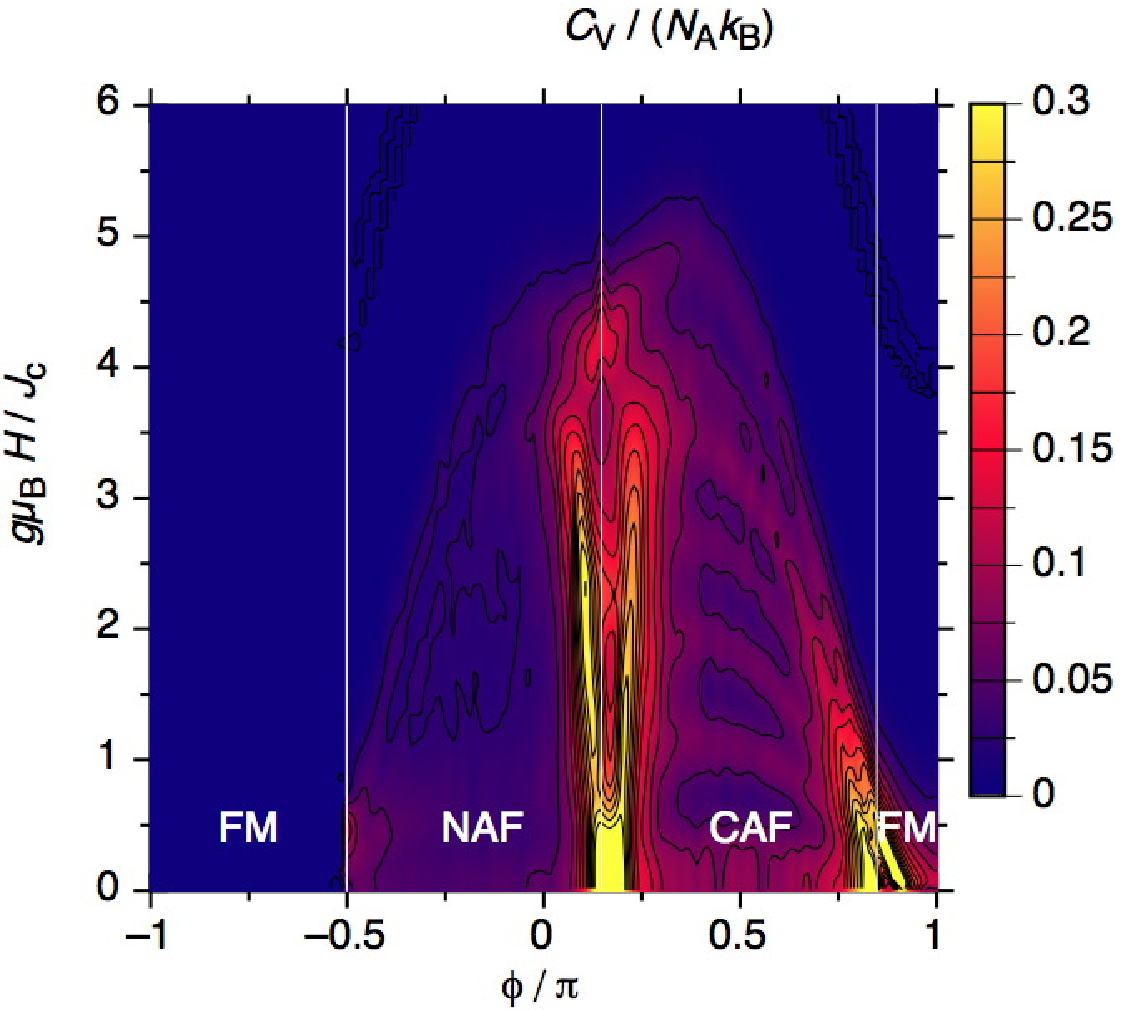}
    \hfill
    \includegraphics[height=.35\textwidth]{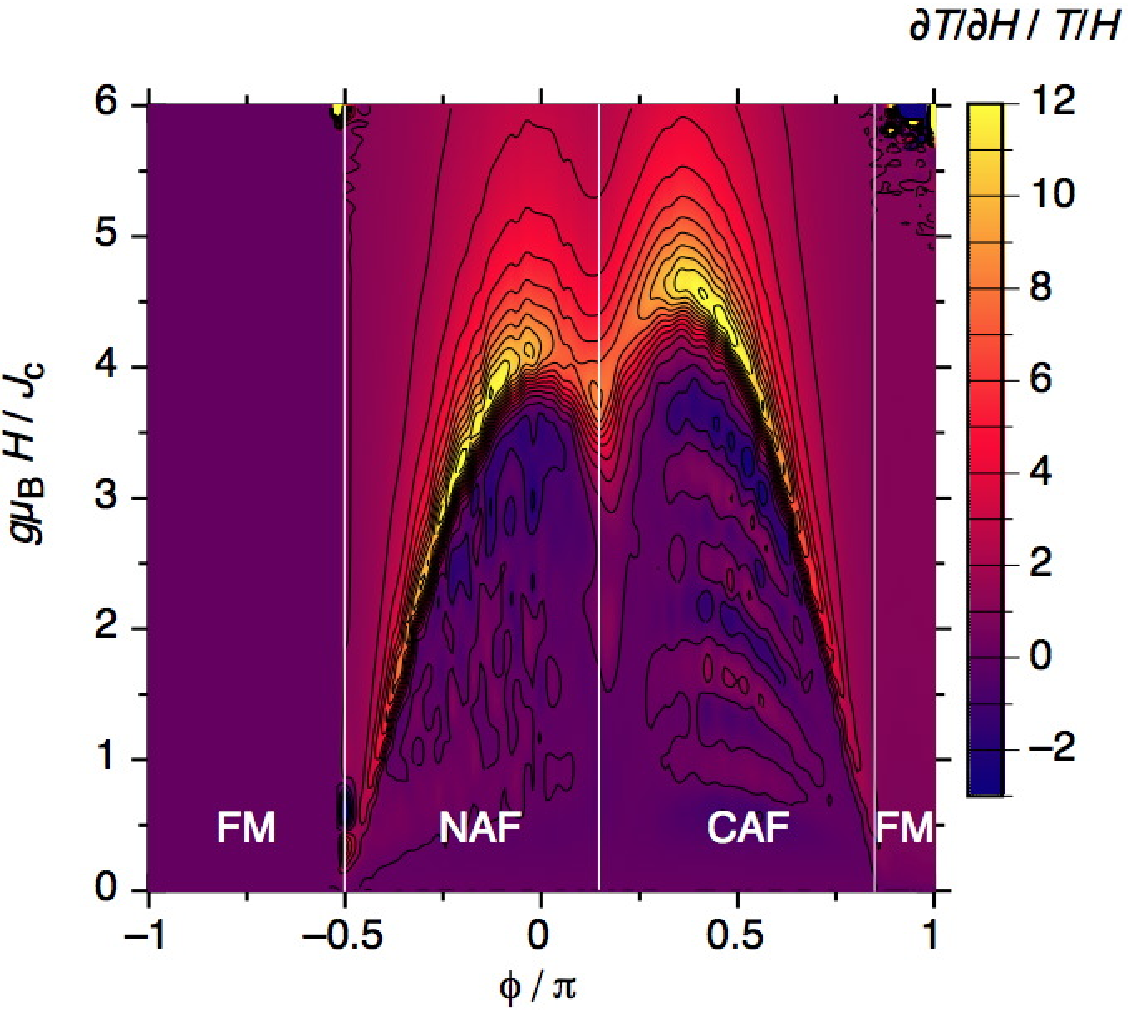}
    \hfill\null \caption{(Color) Contour plots of the heat capacity
    (left) and the normalized magnetocaloric effect
    $\Gamma_{\text{mc}}/(T/H)$ (right) for the 24-site cluster at a
    fixed temperature $T=0.2\,J_{\text c}/k_{\text B}$ as a function
    of the frustration angle $\phi$ and the magnetic field $H$.}
    \label{fig:mceandcv}
\end{figure*}
For the numerical calculation of the magnetocaloric effect we express
Eq.~(\ref{COOLRATE}) as the cumulant
\begin{equation}
    \Gamma_{\text{mc}}\equiv\left(\frac{\partial T}{\partial
H}\right)_{S}=
    -g\mu_{\text B}T
    \frac{\left\langle{\cal H}S_{z}^{\text{tot}}\right\rangle
    -\left\langle{\cal H}\right\rangle
    \left\langle S_{z}^{\text{tot}}\right\rangle}{
    \left\langle{\cal H}^2\right\rangle
    -\left\langle{\cal H}\right\rangle^2}
    \label{eqn:mce}
\end{equation}
and normalize the results to the magnetocaloric effect of a
paramagnet.  The left-hand side of Fig.~\ref{fig:mceandcv} shows a
contour plot of  $\hat{\Gamma}_{\text{mc}}=\Gamma_{\text{mc}}/(T/H)$ as a
function of the
frustration angle $\phi$ and the magnetic field $H$.  For small fields
$H\ll H_{\text{sat}}$, $\Gamma_{\text{mc}}/(T/H)$ is small, nearly
zero or even slightly negative, apart from finite-size effects showing
up in particular in the collinear phase.  It is only at the saturation
field where $\Gamma_{\text{mc}}/(T/H)$ develops a large anomaly peaked
slightly above $H_{\text{sat}}$ (compare with the right plot of
Fig.~\ref{fig:scaling}).  For magnetic fields $H\gg H_{\text{sat}}$,
we eventually reach  $\Gamma_{\text{mc}}/(T/H)\to1$.

\begin{figure}
    \centering
    \includegraphics[width=.35\textwidth]{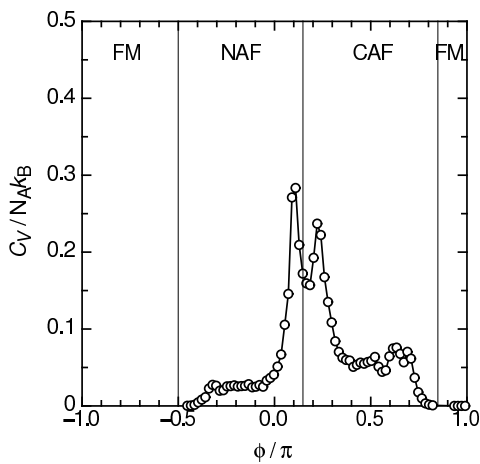}
    \caption{Specific heat at $k_{\text B}T/J_{\text c}=0.2$ and {\em
    constant}, $\phi$-independent field $g\mu_{\text B}H/J_{\text
    c}=2.5$ as function of frustration angle.  The double peak
    structure at the classical NAF/CAF boundary corresponds two the
    two ridges in the contour plot of Fig.~\ref{fig:mceandcv}.}
    \label{fig:cv2peak}
\end{figure}
Apart from a factor $T$, the magnetocaloric effect is given by the
ratio of two quantities, see Eq.~(\ref{COOLRATE}): (a) In the
numerator, we have $(\partial M/\partial T)_{H}$, or, equivalently,
$(\partial S/\partial H)_{T}$.  The entropy $S(T,H)$ at constant
temperature is plotted on the right-hand side of
Fig.~\ref{fig:chiands}.  Its field dependence, corresponding to the
density of contour lines in the plot, is weak apart from the
nonmagnetic regions at the edges of the collinear phase.  (b) The
denominator is the heat capacity $C_{V}(T,H)$.  Here, {\em small\/}
values give rise to a {\em large\/} magnetocaloric effect.
Fig.~\ref{fig:mceandcv} (left) holds a plot of the heat capacity at
constant temperature $T=0.2\,J_{\text c}/k_{\text B}$ as a function of
the frustration angle $\phi$ and the magnetic field $H$.  The heat
capacity is large in the disordered regions, reflecting the high
number of quasi-degenerate states.  Around $J_{2}/J_{1}=1/2$
($\phi/\pi\approx0.148$), a two-peak structure evolves when increasing
the field, see also Fig~\ref{fig:cv2peak}.  Due to the smallness of
the saturation field, we currently cannot say whether such a structure
also exists at the ``mirrored'' ($J_2\rightarrow-J_2$) position in the
phase diagram at $J_{2}/J_{1}=-1/2$.  When reaching the saturation
field, the heat capacity drops and eventually vanishes.

Taken together, it appears naturally that the magnetocaloric effect 
is peaked around the saturation field. The drop 
in magnitude inside the nonmagnetic regions can be understood, too,
as a consequence of their large specific heat.
And since the entropy {\em rises\/} inside these regions when turning 
on the magnetic field, the magnetocaloric effect must be negative, 
indicating a {\em cooling\/} of a sample before reaching the entropy 
maximum.  We note that we also observe a change of sign in 
$(\partial S/\partial H)_{T}$ as a function of field in the 
magnetically ordered regions, and return to this point in the 
context of spin wave theory below.

\begin{figure*}
    \centering
    \hfill
    \includegraphics[height=.35\textwidth]{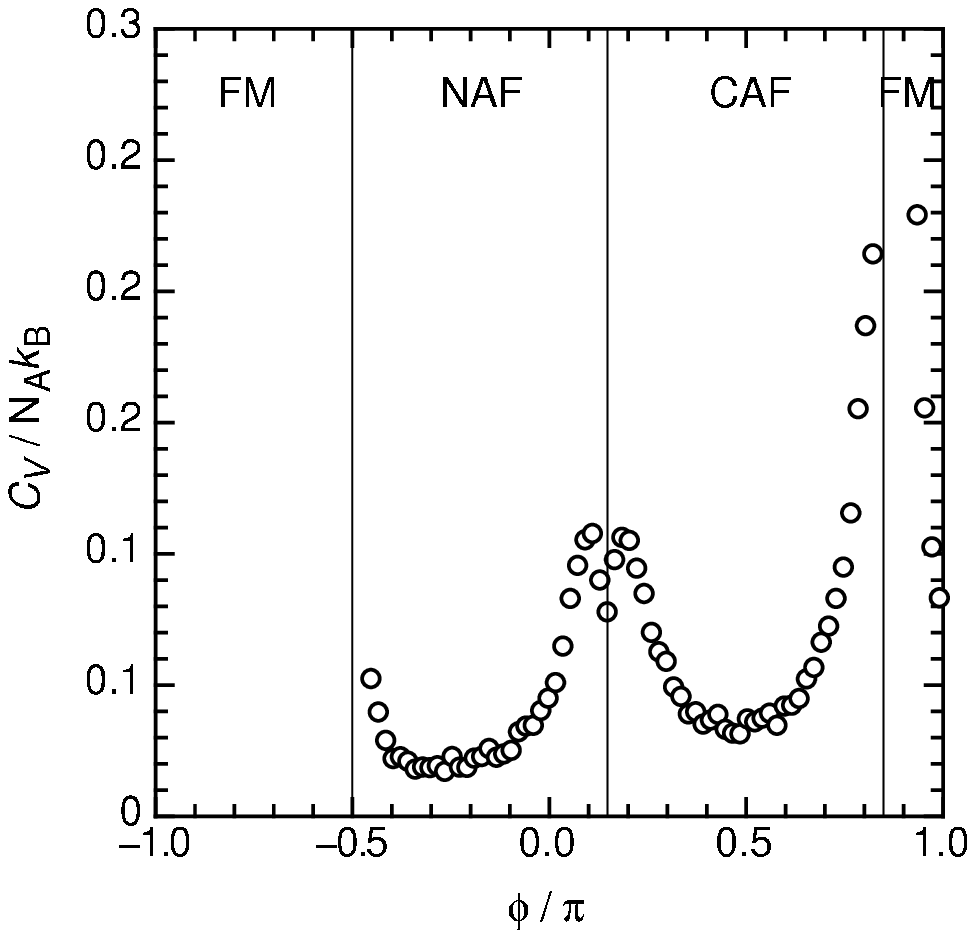}
    \hfill
    \includegraphics[height=.35\textwidth]{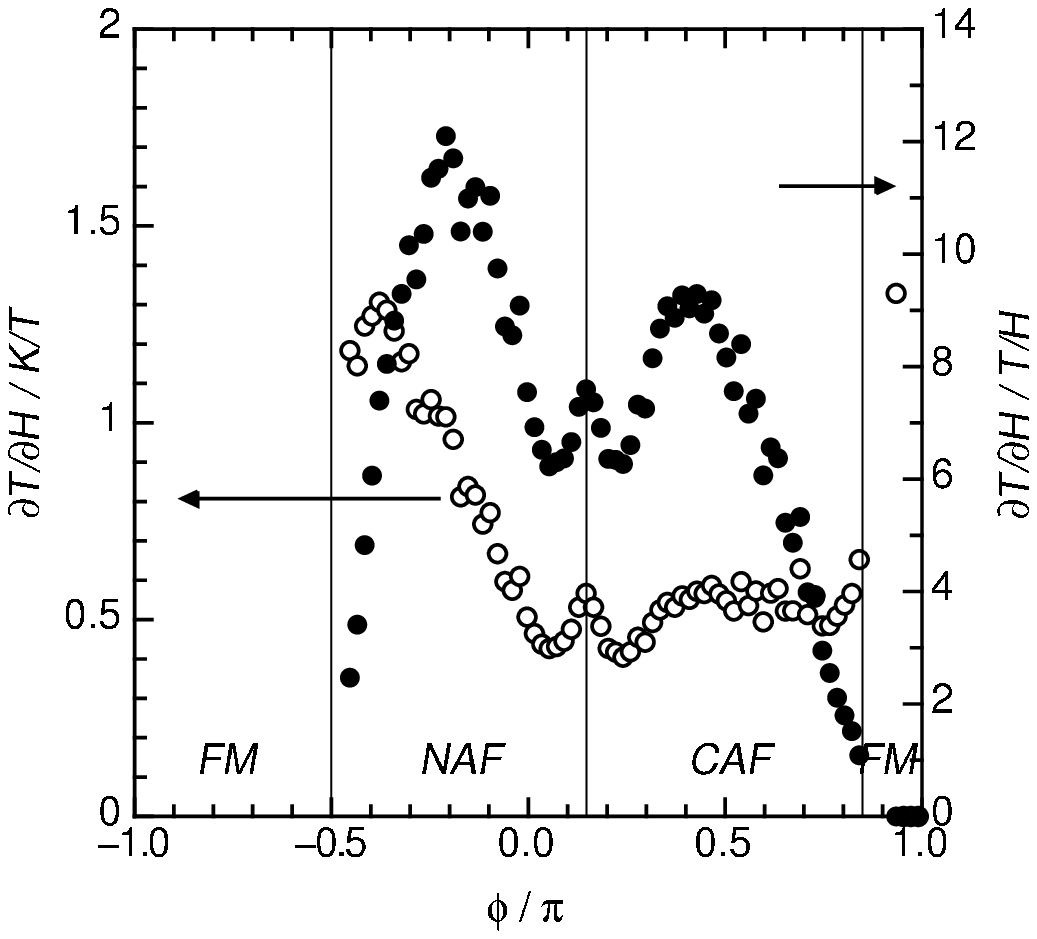}
    \hfill\null \caption{In this figure the field is kept at the
    saturation field for every $\phi$, i.\,e.{} $C_V$ and
    $\Gamma_{\text{mc}}$ are plotted along the curve in
    Fig.~\ref{fig:scaling} (right).  Left: Value of the heat capacity
    $C_{V}(T,H)$ at the saturation field as a function of the
    frustration angle $\phi$ for a fixed temperature $T=0.2\,J_{\text
    c}/k_{\text B}$.  Right: Values of the magnetocaloric effect
    $\Gamma_{\text{mc}}(T,H)=(\partial T/\partial H)_{S}$ at the
    saturation field as a function of the frustration angle $\phi$
    using a fixed temperature $T=0.2\,J_{\text c}/k_{\text B}$.  The
    open circles denote the absolute value of $\Gamma_{\text{mc}}$,
    their scale is given at the left ordinate.  The filled circles
    denote the normalized values $\Gamma_{\text{mc}}/(T/H)$,
    indicating the enhancement relative to a paramagnet.  Scale is on
    the right ordinate.  Both plots were made using a 24-site
    cluster.}
    \label{fig:mceandcvsat}
\end{figure*}
On the right-hand side of Fig.~\ref{fig:mceandcvsat}, we have plotted
the values of $\Gamma_{\text{mc}}(T,H)$ at the saturation field as a
function of the frustration angle $\phi$, again for $T=0.2\,J_{\text
c}/k_{\text B}$.  The open circles denote the absolute values, while
the filled circles denote the values relative to a paramagnet.  In
accordance with the discussion in the previous paragraph, the absolute
values of $\Gamma_{\text{mc}}(T,H)$ in the ordered phases are larger
than in the nonmagnetic regions.  The deviation from the average value
is less than a factor two.  In contrast, the normalization to the
magnetocaloric effect of the paramagnet (filled circles in the right
panel of Fig.~\ref{fig:mceandcvsat}) introduces a strong influence of
the saturation field, compare the right plot in
Fig.~\ref{fig:scaling}.  Therefore, the highest enhancement of
$\Gamma_{\text{mc}}$ with respect to a paramagnet occurs deep inside
the magnetically ordered phases, where the saturation field reaches
its maximum values.

The heat capacity $C_{V}(T,H_{\text{sat}})$ as a function of the
saturation field at constant temperature $T=0.2\,J_{\text c}/k_{\text
B}$ is plotted on the left side of Fig.~\ref{fig:mceandcvsat}.  It is
strongly enhanced in the nonmagnetic regions, while roughly constant
as a function of the frustration angle $\phi$ in the magnetically
ordered phases, giving rise to the comparatively weak $\phi$
dependence of $\Gamma_{\text{mc}}$.  As discussed above, the
enhancement inside the disordered regions is responsible for the
suppression of the magnetocaloric effect.

\begin{figure*}
    \centering
    \hfill
    \includegraphics[width=.35\textwidth]{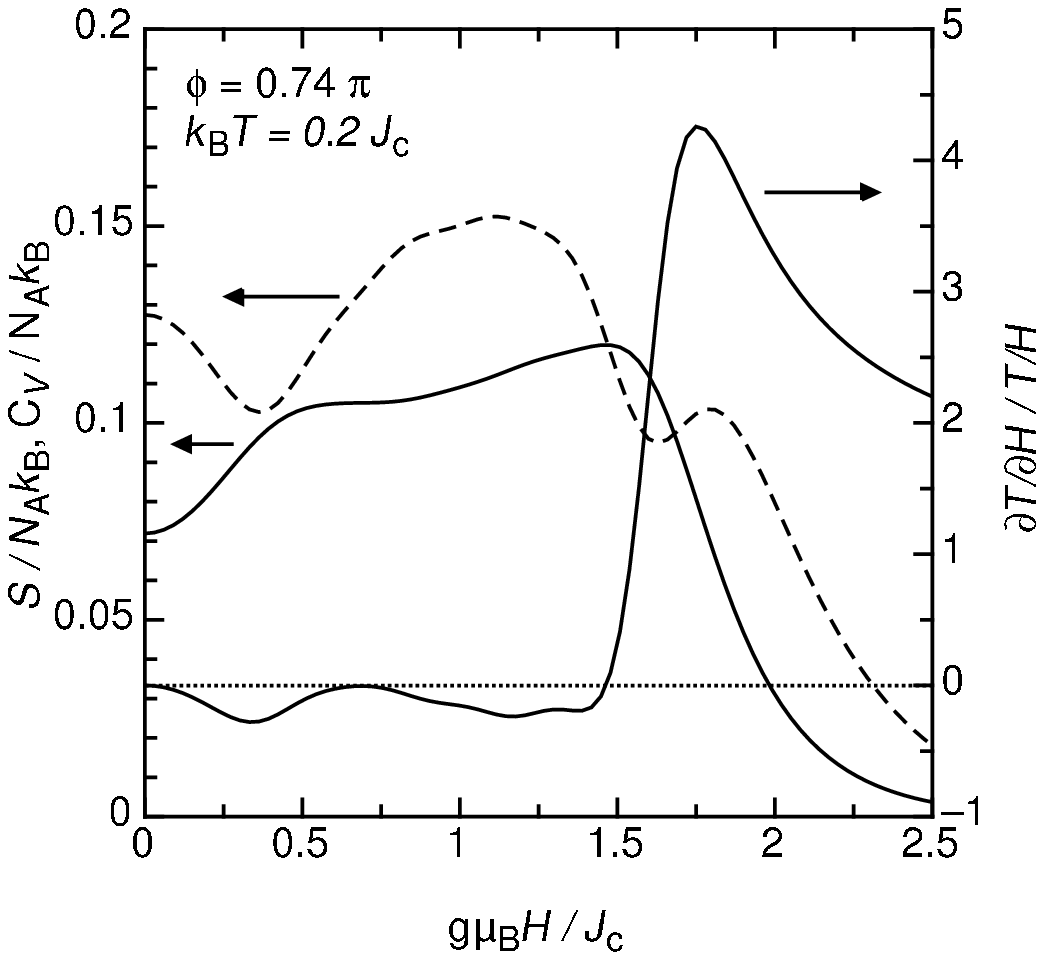}
    \hfill
    \includegraphics[width=.32\textwidth]{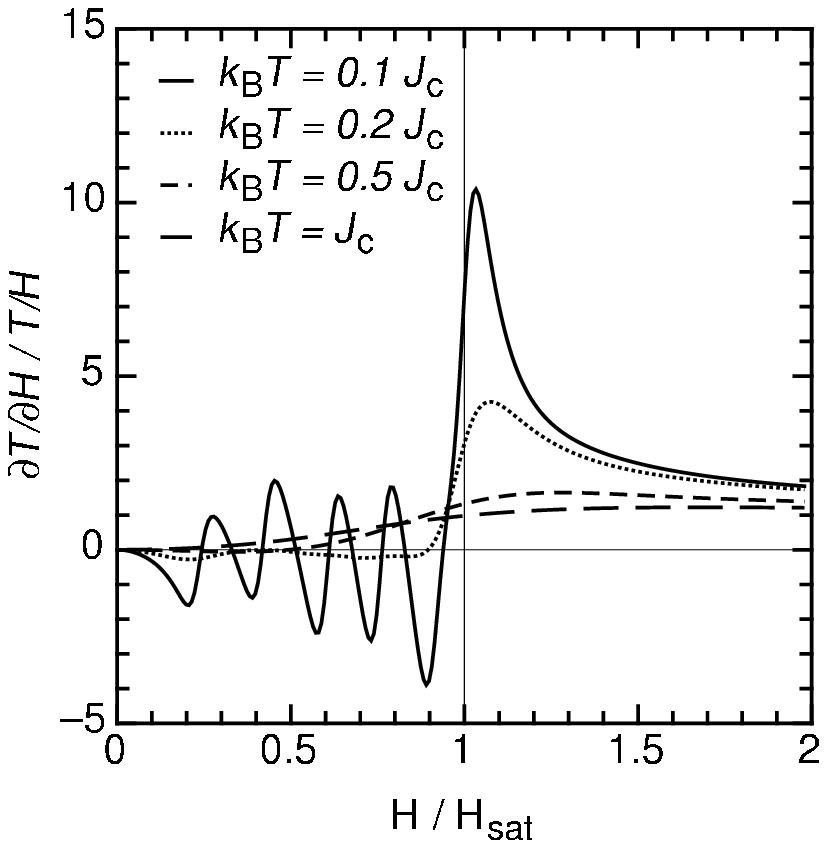}
    \hfill\null\caption{Left: Entropy $S(T,H)$ (solid line, left
    scale), heat capacity $C_{V}(T,H)$ (dashed line, left scale), and
    magnetocaloric effect $(\partial T/\partial H)/(T/H)$ (solid line,
    right scale) as functions of the magnetic field $H$ at constant
    temperature $T=0.2\,J_{\text c}/k_{\text B}$ for a frustration
    angle $\phi=0.74\,\pi$.  Data were generated using a 24-site
    cluster.  Right: Normalized MCE for various temperatures.  The
    anomaly at $H_{\text c}$ is suppressed with increasing $T$, see
    also Fig.~\ref{fig:MCETNAFCAF}.  At the lowest temperature a
    negative MCE is possible.}
    \label{fig:comparison}
\end{figure*}

To clarify the field dependence of $\Gamma_{\text{mc}}(T,H)$ further,
Fig.~\ref{fig:comparison} holds a comparison of the three relevant
quantities $\Gamma_{\text{mc}}(T,H)/(T/H)=(\partial T/\partial
H)/(T/H)$, $C_{V}(T,H)$, and $S(T,H)$ as a function of the magnetic
field $H$ for a fixed frustration angle $\phi=0.747\,\pi$ and a
temperature $T=0.2\,J_{\text c}/k_{\text B}$: Disregarding possible
finite-size effects, the entropy (solid line, left scale) and the heat
capacity (dotted line, left scale) are slowly varying functions of
$H$, see also Sec.~\ref{sec:MF}, dropping sharply above the
saturation field $H_{\text
{sat}}\approx1.64\,J_{\text c}/(g\mu_{\text B})$.  Taken together,
this leads to a pronounced maximum of $(\partial T/\partial H)/(T/H)$
slightly above the saturation field.  Otherwise, $(\partial T/\partial
H)/(T/H)$ is small and negative for fields $H\ll H_{\text{sat}}$
(because $(\partial S/\partial H)_{T}\ge0$ in this field range) and
approaches $1$ for $H\gg H_{\text{sat}}$.

\section{Approximate analytical treatments of the model}

\label{sect:ANALYTIC}

To better understand the exact numerical results for finite clusters
it is useful to have approximate analytical results available for
comparison.  We consider two approaches: Firstly a mean field
treatment which provides a reference point for the global behavior of
entropy and specific heat in the ordered phase, and the magnetocaloric
effect above the saturation field.  Secondly we use a linear spin-wave
(LSW) approximation to investigate the anomalous enhancement of the
magnetocaloric effect around the saturation field, which turns out to
be due to a softening of spin excitations at characteristic wave
vectors.  In this approximation, in contrast to mean field theory, the
MCE below the saturation field is nonzero.  The spin wave
approximation also allows to study subtle effects for subcritical
fields which lead to a sign change of the MCE.

The existence of long range order at finite temperatures, i.\,e.{} a
non-vanishing transition temperature $T_{\text c}$ associated with
magnetic order, is implicit in both treatments.  In reality, for the
layered vanadates which we wish to describe, $T_{\text c}$ will be
determined by low energy scales which are not present in our model,
notably the interlayer magnetic exchange $J_\perp$ and magnetic
anisotropy~$\delta J$.

Formally, we cannot break a continuous symmetry such as spin rotation
at {\em any\/} finite temperature in 2D, and to be ``correct'' we
should generalize our model to higher dimension and finite anisotropy.
However these details make little qualitative (or quantitative)
difference for a wide range of temperatures $\delta J, J_\perp \ll T
\ll T_{\text c}$, so we suppress them below.  Furthermore, in these
calculations we neglect the consequence of interactions between spin
waves (see e.g. \cite{jackeli:04}) and the break-down of magnetic
order at {\em zero\/} temperature on the borders of the CAF
phase~\cite{sindzingre:04,shannon:06}.  These effects can be expected
to modify the details of critical behavior as a function of magnetic
field, but not its broad features, and are left for future
investigation.

As discussed in the previous section, the size of the magnetocaloric
cooling rate in Eq.~(\ref{COOLRATE}) is determined by the ratio of the
rate of change of entropy with field $\left(\partial S/\partial
H\right)_{T,V}$ and its rate of change with temperature
$\left(\partial S/\partial T\right)_{H,V} =C_V/T$.  On approaching the
critical field both quantities tend to increase sharply, and the
resulting increase in $\Gamma_{\text{mc}}(H)$ is a tradeoff between
them.  It is not immediately obvious for which frustration angle
$\phi$ the enhancement in $\Gamma_{\text{mc}}(H_{\text c},\phi)$
should be largest.  At modest temperatures, the simple spin wave
approximation described below gives considerable insight into this
question, and the critical anomalies of the MCE around the saturation
field.

\subsection{Calculation of mean field order parameters and
thermodynamics}
\label{sec:MF}
In this section the magnetothermal properties will be investigated in
mean field approximation to have a reference for the spin wave and
numerical exact-diagonalization methods.  The results of the former
are, however, not expected to give a realistic description of the MCE.
For a unified treatment of AF phases it is advisable to use a
four-sublattice description ($\alpha,\beta = \text A,\text B,\text
C,\text D$) with each sublattice having $N/4$ sites for both NAF and
CAF. Since we consider only isotropic exchange we may assume without
loss of generality that the field is perpendicular to the (xy) plane
of the square lattice, i.\,e.{}, $\bh=h\hat{\bz}$.

In this and the following subsection we refer all extensive quantities
like entropy, specific heat etc.{} to a single site for convenience.
The exchange field $\bh^{\text{ex}}$ and total molecular field
$\hat{\bh}$ due to Eq.~(\ref{HAMJ1J2}) is then given by
\begin{eqnarray}
\label{EXFIELD1}
\bh_\alpha^{\text{ex}}&=&-\sum_{k\beta}J_{\alpha\beta}^{lk}\la\bS_\beta\ra\no\\
\hat{\bh}_\alpha&=&\bh+\bh_\alpha^{\text{ex}}
\end{eqnarray}
where the exchange constants $J_{\alpha\beta}^{lk}$ are defined per
bond.  The components of the exchange field $h_\parallel^{\text{ex}}$
and $h_\perp^{\text{ex}}$ which are parallel and perpendicular to the
field direction z are related to the respective spin expectation
values $\la S_\parallel\ra$ and $\la S_\perp\ra$ via the equations
\begin{eqnarray}
\label{EXFIELD2}
\frac{1}{2}h_\parallel^{\text{ex}}=-a_\parallel\la S_\parallel\ra;\quad
 \frac{1}{2}h_\perp^{\text{ex}}=a_\perp\la S_\perp\ra
\end{eqnarray}
where the prefactors for the AF and the FM or fully polarized  phases
($h>h_{\text c}$ for any $\phi$) are given by
\begin{eqnarray}
\label{ACOEFF}
\mbox{NAF:}\quad a_\parallel&=&\frac{z}{2}(J_1+J_2);\quad
a_\perp=\frac{z}{2}(J_1-J_2)\no\\
\mbox{CAF:}\quad a_\parallel&=&\frac{z}{2}(J_1+J_2);\quad
a_\perp=\frac{z}{2}J_2\\
\mbox{FM:}\quad a_\parallel&=&\frac{z}{2}(J_1+J_2);\quad a_\perp=0\no
\end{eqnarray}

Then the mean field approximation of the Hamiltonian in
Eq.~(\ref{HAMJ1J2}) may be written as
\begin{eqnarray}
\label{HMF}
H_{\text{mf}}=\sum_{\alpha,l}[-(\bh+\bh_\alpha^{\text{ex}})\bS_\alpha^l
+\frac{1}{2}\bh_\alpha^{\text{ex}}\la\bS_\alpha\ra]
\end{eqnarray}
This also defines selfconsistently the mean field averages via $\la
A\ra = \mathop{\rm Tr}\{A\exp(-\beta H_{\text{mf}})\}/\mathop{\rm
Tr}\{\exp(-\beta H_{\text{mf}})\}$ with $\beta=1/(k_{\text B}T)$.
From the above equation the corresponding total mean field internal
energy per site $U_{\text{mf}}=(1/N)\la H_{\text{mf}}\ra$ is obtained
as
\begin{eqnarray}
    \label{UEN1}
    U_{\text{mf}}(T,H)
    =-\frac{1}{4}\sum_{\alpha}(\bh+\frac{1}{2}\bh_\alpha^{\text{ex}})
    \la\bS_\alpha\ra
\end{eqnarray}
Explicitly, using Eq.~(\ref{ACOEFF}) one obtains 
\begin{eqnarray}
    \label{UEN2}
    \mbox{NAF, CAF:}\quad U_{\text{mf}}&=&-h\la S_\parallel\ra+a_\parallel\la
    S_\parallel\ra^2-a_\perp\la S_\perp\ra^2\no\\
    \mbox{FM:}\quad U_{\text{mf}}&=&-h\la S_\parallel\ra+a_\parallel\la
    S_\parallel\ra^2
\end{eqnarray}

\begin{figure*}
    \vspace{5mm}
    \centering
    \includegraphics[width=0.5\textwidth]{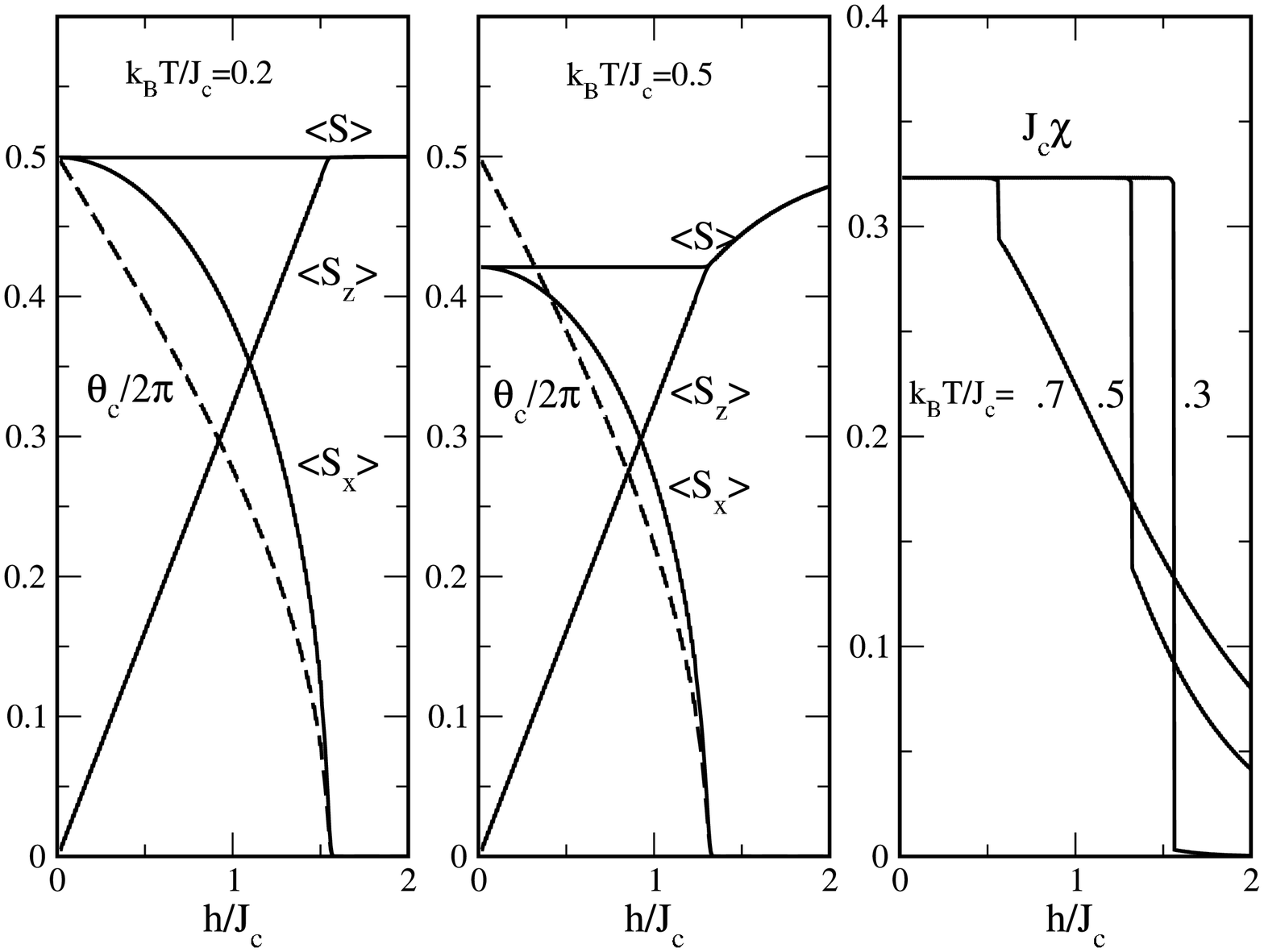}
    \caption{Field dependence of staggered order parameter $\la
    S_x\ra$ and uniform magnetization $\la S_z\ra$ for two different
    temperatures (left and center panel).  The total moment $\la
    S\ra=(\la S_x\ra^2+\la S_z\ra^2)^\frac{1}{2}$ is field independent
    in the ordered regime.  The field dependence of the canting angle
    $\theta_{\text c}/2$ (counted from the field direction) is also
    shown.  Right panel: Susceptibility $\chi$ for three different
    temperatures.  Its value in the ordered regime is $T$-independent.
    In all cases the frustration angle is $\phi=0.74\pi$ corresponding
    to the CAF choice.}
    \label{fig:OPh}
\end{figure*}
To calculate thermodynamic quantities the expectation values $\la
S_\parallel\ra$ and $\la S_\perp\ra$ and their temperature derivatives
have to be obtained selfconsistently.  This is done by diagonalizing
$H_{\text{mf}}$ which leads to local eigenstates
$|\pm\ra=u_\pm|\uparrow{}\ra+v_\pm|\downarrow{}\ra$ where
$|\uparrow{}\ra$, $|\downarrow{}\ra$ are the degenerate free $S=1/2$
states.  The $|\pm\ra$ states have energies $E_\pm=E_{\text
c}+\epsilon_\pm$ given by
\begin{eqnarray}
    \label{EIGENE}
    E_{\text c}&=&a_\pe\la S_\pe\ra^2-a_\pa\la S_\pa\ra^2\no\\
    \epsilon_\pm&=&\pm\frac{1}{2}[\hh_\pa^2+\hh_\pe^2]^\frac{1}{2}
    \nonumber\\
    &&{}\mbox{with}\quad
    \hh_\pa=h+h^{\text{ex}}_\pa;\quad \hh_\pe=h^{\text{ex}}_\pe.
\end{eqnarray}
Defining $2\epsilon_0=\Delta=\epsilon_+-\epsilon_-$ as the splitting
due to the molecular field $\hat{\bh}$  their coefficients are then
obtained as
\begin{eqnarray}
    \label{EIGENCO}
    u_\pm&=&\frac{\frac{1}{2}\hh_\pe}%
    {[\epsilon_0(2\epsilon_0\pm\hh_\pa)]^\frac{1}{2}};
    \quad
    v_\pm=\mp\frac{\frac{1}{2} (2\epsilon_0\pm\hh_\pa)^\frac{1}{2}
    }{\epsilon_0^\frac{1}{2}}
\end{eqnarray}

Using these coefficients and the difference of thermal occupation
numbers $p_--p_+=\tanh\frac{1}{2}\beta\Delta$ of eigenstates $|\pm\ra$
one finally obtains selfconsistent equations
for the spin expectation values:
\begin{eqnarray}
\label{SELFCONS}
\la S_\pe\ra&=&\frac{1}{2}\frac{\hh_\pe}{\Delta}\tanh\frac{1}{2}\beta\Delta,
\no\\
\la S_\pa\ra&=&\frac{1}{2}\frac{\hh_\pa}{\Delta}\tanh\frac{1}{2}\beta\Delta
\end{eqnarray}
The selfconsistency is implied via  the molecular field expressions
\begin{eqnarray}
\label{MOLFIELD}
\hh_\pa&=&h-2a_\pa\la S_\pa\ra\no\\
\hh_\pe&=&2a_\pe\la S_\pe\ra\\
\Delta&=&(\hh_\pa^2+\hh_\pe^2)^\frac{1}{2}\no
\end{eqnarray}
The canting angle $\theta_{\text c}/2$ of magnetic moments is obtained
from minimizing $U_{\text{mf}}$ in Eq.~(\ref{UEN2}) The angle is
counted from the field- or c-direction and given by
\begin{eqnarray}
\tan(\frac{\theta_{\text c}}{2})&=&\frac{\la S_x\ra}{\la
S_z\ra}\quad\mbox{or}
\\
\cos(\frac{\theta_{\text c}}{2})&=&\frac{h}{2\la S\ra}\frac{1}{a_\pa+a_\pe}
=\frac{h}{h_c}
\label{MFANGLE}
\end{eqnarray}
where h$_{\text c}$=2$\langle {\text S}\rangle$(a$_\parallel$ +a$_\perp$).
Inserting this into Eqs.~(\ref{SELFCONS},\ref{MOLFIELD}) leads to the
simple and general result
\begin{eqnarray}
\la S\ra=\frac{1}{2}\tanh\frac{1}{2}\beta\Delta
\quad\mbox{with}\quad
\Delta= 2\la S\ra|a_\perp|
\label{ROTATE}
\end{eqnarray}
This means that in the ordered phase the molecular field splitting
$\Delta$ of spins is field independent up to $h_{\text c}$ and hence
the total moment $\la S\ra$ is also field independent, i.\,e.{} the
moment can only be rotated by the field as long as the transverse
staggered order exists.  This fact has striking consequences for the
thermodynamic quantities below $h_{\text c}$.  For the thermodynamics
we also need the temperature derivatives $\partial\la S_i\ra/\partial
T=-k_{\text B}\beta^2(\partial \la S_i\ra/\partial \beta)$
($i=\pa,\pe$).  They are obtained from Eq.~(\ref{SELFCONS}) in a
straightforward but lengthy calculation and the resulting explicit
expressions are given in appendix A.

\begin{figure*}
    \vspace{5mm}
    \centering
    \includegraphics[width=0.5\textwidth]{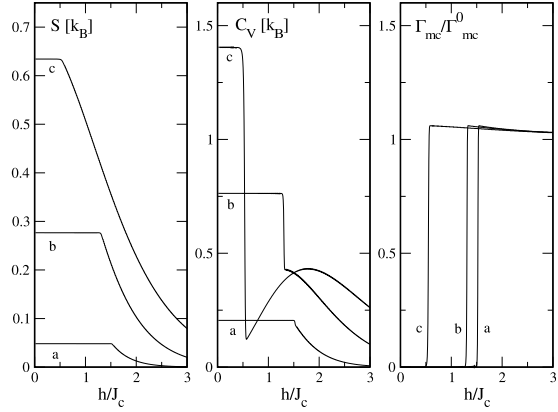}
    \caption{Thermodynamic properties in mean field approximation for
    $\phi=0.74\pi$ (CAF) as function of applied field for three
    subcritical temperatures $k_{\text B}T/J_{\text c}=0.3$ (a), $0.5$
    (b) and $0.7$ (c).  Left: Entropy dependence on $h$.  Below the
    saturation field $h_{\text c}$ it is field independent because
    $\la S\ra$ (see Fig.~\protect\ref{fig:OPh}) and also $\Delta$ is
    field independent in the ordered regime.  Center: specific heat
    dependence on $h$.  For larger temperatures when $h_{\text c}$ is
    sufficiently suppressed a Schottky peak evolves above $h_{\text
    c}$.  Right: Magnetocaloric enhancement factor.  In MF
    approximation almost no enhancement is visible above $h_{\text
    c}(T)$.  Below the critical field the entropy is field independent
    and hence $\Gamma_{\text{mc}}$ drops to zero suddenly according to
    Eq.~(\protect\ref{COOLRATE}).}
    \label{fig:MFTHERMH}
\end{figure*}
The mean field solution for a CAF value of $\phi = 0.74\,\pi$ is shown
in Fig.~\ref{fig:OPh}.  On the left panel the decrease of the
staggered OP $\la S_x\ra$ with increasing field and the concomitant
increase of the uniform moment $\la S_z\ra$ are shown for small
temperature.  The total moment $\la S\ra$ is constant as predicted and
practically equal to S=1/2 in the whole field range.  This confirms
that the moment is simply rotated (canted) by the field without
changing its size.  The relevant canting angle $\theta_{\text c}/2$ is
also shown in the figure.  For moderate temperatures (center panel)
the zero field value of $\la S_x\ra$ is already reduced somewhat.  As
required by Eq.~(\ref{ROTATE}) the field still only reorients the
moment, i.\,e., $\la S\ra$ is a constant less than $1/2$ for fields
$h<h_{\text c}$.  Finally for $h>h_{\text c}$ when the moment is
aligned with the field ($\theta_{\text c}/2=0$), the total moment $\la
S\ra=\la S_z\ra$ will be polarized, i.\,e.{} increases with field
until it reaches the asymptotic value of $S=1/2$.  The right panel
shows the susceptibility $\chi_{\text{mf}}$ for various temperatures.
It is constant and $T$-independent in the ordered regime.  These
results are qualitatively unchanged for different angles $\phi$.

The desired thermodynamic quantities may now be conveniently obtained
from the free energy $F_{\text{mf}}$ and internal energy $U_{\text{mf}}$
(Eq.~\ref{UEN2}) of the $S=1/2$ system split by the molecular field by
an energy $\Delta(\hh_\pa,\hh_\pe)$. The former is  given by
\begin{equation}
    \label{FUEN}
    F_{\text{mf}}(T,H)=E_{\text
    c}-\frac{1}{\beta}\ln[2\cosh\frac{1}{2}(\beta\Delta)]
\end{equation}
per site and the entropy $S_{\text{mf}}=-(\partial
F_{\text{mf}}/\partial T)$, specific heat $C^{\text{mf}}_V=(\partial
U_{\text{mf}}/\partial T)$ and susceptibility per site are then
obtained as
\begin{eqnarray}
    \label{ENTMF}
    S_{\text{mf}}&=&k_{\text B}[\ln(2\cosh\frac{1}{2}\beta\Delta)
    -\frac{1}{2}\beta\Delta\tanh\frac{1}{2}\beta\Delta] \no\\
         C^{\text{mf}}_V&=&k_{\text B}\beta^2[(h-2a_\pa\la S_\pa\ra)
	 \la S_\pa\ra'+2a_\pe\la
    S_\pe\ra']\no\\
    \chi_{\text{mf}}&=&(g\mu_{\text B})^2\partial\la S_\pa\ra/\partial h\no
\end{eqnarray}
where $\la S_\pa\ra'$,  $\la S_\pe\ra'$ are given in
Eq.~(\ref{STGRAD}) of Appendix~A.
For uncoupled spins $a_\pa=a_\pe=0$ and $\la S_\pa\ra'=
\frac{1}{4}h\cosh^{-2}\frac{1}{2}\beta h$ which leads to the Schottky
specific heat of the two level system.  For the magnetocaloric cooling
rate $\Gamma_{\text{mc}}$ we need in addition the temperature gradient
of the magnetization $m_{\text{mf}}=g\mu_{\text B}\la S_\pa\ra$ which
is simply given by
\begin{eqnarray}
\label{MGRADMF}
\frac{\partial m_{\text{mf}}}{\partial T}=-k_{\text B}(g\mu_{\text
B})\beta^2\la S_\pa\ra'
\end{eqnarray}
Then the mean field expression for the cooling rate
$\Gamma_{\text{mc}}$ may be obtained from the definition in
Eq.~(\ref{COOLRATE}) using the expressions for $C_V$ in
Eq.~(\ref{ENTMF}) and the temperature gradient in Eq.~(\ref{MGRADMF}).
Thus the solution of the selfconsistent Eqs.~(\ref{SELFCONS}) for $\la
S_\parallel\ra$ and $\la S_\perp\ra$ and their temperature gradients
$\la S_\parallel\ra'$ and $\la S_\perp\ra'$ in Eq.~(\ref{STGRAD})
provide all the necessary input for obtaining the magnetocaloric
quantities $S_{\text{mf}}(T,H)$, $C^{\text{mf}}_V(T,H)$ and
$\Gamma_{\text{mc}}(T,H)$ from the above equations.

In Fig.~\ref{fig:MFTHERMH} we show the field dependence of mean field
entropy, specific heat and cooling rate as function of field for
various temperatures.  Obviously $S$ and $C_V$ (left and center panel)
are field independent below $h_{\text c}$ caused by the fact that
$\Delta$ is constant according to Eq.~(\ref{ROTATE}).  Consequently,
since the cooling rate is proportional to the field-gradient of $S$
(Eq.~\ref{COOLRATE}), it suddenly drops to zero below the saturation
field $h_{\text c}$ as seen in the right panel of
Fig.~\ref{fig:MFTHERMH}.  In addition it shows that above $h_{\text
c}$ the cooling rate is only slightly enhanced from the paramagnetic
value.  Although the mean field results are far from realistic, the
two main aspects, field independence of $S$ and $C_V$ below $h_{\text
c}$ and steplike anomaly in $\Gamma_{\text{mc}}$ at $h_{\text c}$
still leave their signature in the more advanced spin wave and
numerical treatment to be discussed below.

\subsection{The magnetocaloric effect in the linear spin wave
approximation}

\subsubsection{Statistical mechanics and general expressions}
\label{onespin}

In mean field approximation the elementary excitations are local
dispersionless spin flips whose energy is determined by the molecular
field.  This is far from reality especially close to the saturation
field when the order parameter breaks down, associated with a
softening of the spin excitations at some $\bk$-point or even line in
the Brillouin zone (BZ).  The linear spin wave approximation takes
this effect into account and gives a much more realistic description
of the MCE. Since spin wave interactions are left out in our approach,
however, the singular behavior around $h_{\text c}$ may be
overestimated.

\begin{figure*}
    \centering
    \hfill
    \includegraphics[width=0.35\textwidth]{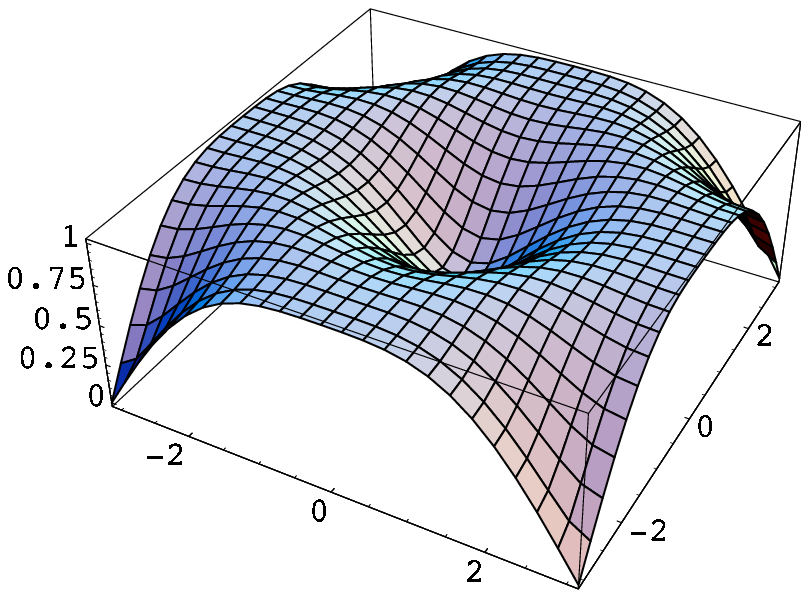}
    \hfill
    \includegraphics[width=0.35\textwidth]{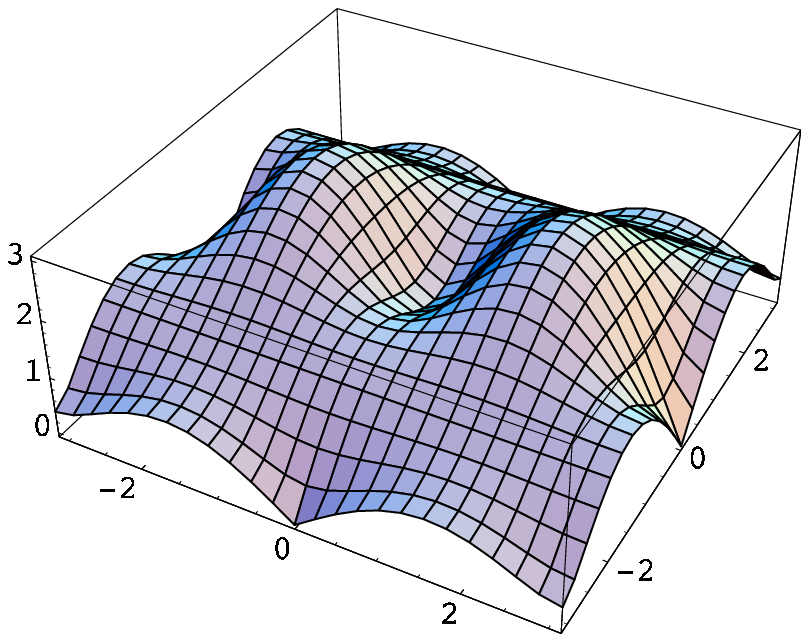}
    \hfill\null\caption{(Color) Spin wave dispersions in the NAF and
    CAF phase.  Only the $\epsilon_+ (h,{\bf k})$ modes are shown.
    The corresponding $\epsilon_- (h,{\bf k})$ modes are obtained by
    translation with $(\pi,\pi)$ or $(\pi,0)$ in the NAF and CAF case
    respectively.  Left panel: canted NAF dispersion $\epsilon_+
    (h,{\bf k})$ for $J_2=0$ and a canting angle of $\theta_{\text c}
    = 7\pi/8$.  Note that $\epsilon_+ (h,{\bf k})$ is gapless at
    ($\pi,\pi$) and gapped at (0,0) while the opposite holds for
    $\epsilon_- (h,{\bf k})$.  Right panel: canted CAF dispersion
    $\epsilon_+ (h,{\bf k})$ for $J_1=1$, $J_2=1$, $S=1/2$ and a
    canting angle of $\theta_{\text c} = 7\pi/8$.  In this case
    $\epsilon_+ (h,{\bf k})$ is gapless at ($\pi,0$) and gapped at
    (0,0) and vice versa for $\epsilon_- (h,{\bf k})$.  }
    \label{fig:omega1NAFcanted}
    \vspace{0.5cm}
\end{figure*}
To calculate the magnetocaloric effect of
Eq.~(\ref{COOLRATE}) in spin wave approximation we start from the
partition function
\begin{eqnarray}
    Z = {\rm Tr} [e^{-{\cal H}/k_{\text B}T} ]
\end{eqnarray}    
where the Hamiltonian $\cal{H}$ is expanded in spin wave
coordinates using the  Holstein-Primakoff approximation
\begin{eqnarray}
    {\cal H} &=& NE_0  +NE_{\text{zp}} + \sum_{\lambda{\bf k}} 
    \epsilon_\lambda(h,{\bf   k})\alpha^{\dagger}_{\lambda{\bf k}} 
    \alpha^{\phantom\dagger}_{\lambda{\bf k}}
    \nonumber\\&&{}
    + {\cal O}(E_0/S^2)
    \label{HPHAM}
\end{eqnarray}   
Here $E_0=U_{\text{mf}}(T=0)$ is the classical (mean field) ground
state energy per spin, $\epsilon_\lambda(h,{\bf   k})$ the spin 
wave  dispersion in applied field $h$, and the sum
over ${\bf k}$ runs over the appropriate magnetic BZ, while $\lambda$
counts the different spin wave branches within that BZ. 
The operator
$\alpha^{\dagger}_{\lambda{\bf k}}$ creates magnons with
commutation relations 
\begin{eqnarray}
[ \alpha^{\phantom\dagger}_{\lambda{\bf k}}
, \alpha^{\dagger}_{\lambda'{\bf k'}}
]=\delta_{\lambda\lambda'}\delta_{\bf kk'}\nonumber. 
\end{eqnarray}  
In addition to the classical (mean field) ground
state energy per spin $E_0$ there is zero point 
energy contribution 
\begin{eqnarray}
 E_{\text{zp}}=\frac{1}{2N}\sum_{\lambda {\bf k}}\left[ \epsilon_\lambda(h,{\bf 
 k}) - A(h,{\bf k})\right] 
\end{eqnarray}   
where $A(h,{\bf k})$ is the on-sublattice coupling between spins,
defined below.  For CAF and NAF phases the E$_{\text{zp}}$ is always
negative; in the FM, where the ground state and spin waves are
eigenstates with a single dispersion $\epsilon(h,{\bf k}) \equiv
A(h,{\bf k})$, E$_{\text{zp}}$ vanishes identically.

The partition function is essentially that of set of independent 
simple harmonic oscillators
\begin{eqnarray}
    Z &=& 
    e^{-[E_0(h)+E_{\text{zp}}(h)]/(k_{\text B}T)}\no\\ 
    &&\times\prod_{\lambda{\bf k}}
    \left[1 - e^{-\epsilon_\lambda(h,{\bf k})/(k_{\text B}T)}\right]^{-1}
\end{eqnarray}
From this we find the internal and free energy per site, using $n_{\text
B}(\epsilon, T) = [e^{\epsilon/(k_{\text B}T)} - 1]^{-1}$ for the Bose
factor:
\begin{eqnarray}
    U&=&E_0+\frac{1}{N}\sum_{\lambda,\bk}n_{\text B}[\epsilon_\lambda(h,{\bf k})]
    \epsilon_{\lambda}(h,{\bf k})\no\\
    F&=& -\frac{1}{N}\,k_{\text B}T\ln Z
    \\
    &=& E_0(h) + E_{\text{zp}}(h)
    \nonumber\\&&{}
    + \frac{1}{N}\;k_{\text B}T\sum_{\lambda{\bf k}} \ln \left[1 -
    e^{-\epsilon_{\lambda}(h,{\bf k})/(k_{\text B}T)}\right]
    \nonumber
\end{eqnarray} 
The entropy per site $S=-(\partial F/\partial T)$ follows directly :
\begin{eqnarray}
S=\frac{k_{\text B}}{N}\sum_{\lambda,\bk}\Bigl[\frac{1}{2k_{\text 
B}T}\mathop{\rm ctnh}\frac{\epsilon_\lambda(h,{\bf
k})}{2k_{\text B}T}
-\ln\sinh\frac{\epsilon_\lambda(h,{\bf k})}{2k_{\text B}T}\Bigr]
\label{SLSW}
\end{eqnarray}
We can also find the uniform magnetization 
\begin{eqnarray}
    m &=& -\frac{\partial E_0(h)}{\partial h} 
    -\frac{\partial E_{\text{zp}}(h)}{\partial h}
    \nonumber\\
    &&{}
    - \frac{1}{N}\sum_{\lambda{\bf k}} 
    n_{\text B}[\epsilon_{\lambda}(h,{\bf k}),T]
    \frac{\partial \epsilon_\lambda(h,{\bf k})}{\partial h} 
    \label{MLSW}
\end{eqnarray}
In the FM phase, where $\epsilon(h,{\bf k}) = \omega_k + h$ and there
is no zero--point term, this simply reduces to 
\begin{eqnarray}
m = m_0 - \frac{1}{N}\sum_{\lambda{\bf k}} n_{\text B}(\omega_k + h, T)
\end{eqnarray}
with $m_0=\partial E_0(h)/\partial h$.  

Quite generally, we can calculate the MCE
as the ratio in Eq.~(\ref{COOLRATE}) where the magnetization gradient
is given by
\begin{eqnarray}
\label{MGRADLSW}
 \frac{\partial m}{\partial T} = 
 -\frac{1}{N}\sum_{\lambda{\bf k}} \frac{\epsilon_\lambda(h,{\bf k}) 
 \frac{\partial \epsilon_\lambda(h,{\bf k})}{\partial h}}
 {4 (k_{\text B}T)^2 \sinh^2\left[\frac{\epsilon_\lambda(h,{\bf k})}{2
k_{\text B}T}\right]}
\end{eqnarray}
and the specific heat $C_V=T(\partial S/\partial T)=(\partial
U/\partial T)$ by
\begin{eqnarray}
\label{SPECLSW}
 \frac{C_V}{T} = 
 \frac{k_{\text B}}{N}\sum_{\lambda{\bf k}} \frac{\epsilon_\lambda
 (h,{\bf k})^2}{4 (k_{\text B}T)^3 \sinh^2\left[\frac{\epsilon_\lambda(h,{\bf
k})}{2 k_{\text B}T}\right]}
\end{eqnarray} 
From Eqs.~(\ref{MGRADLSW}, \ref{SPECLSW}), $\Gamma_{\text{mc}}$ is
obtained using Eq.~(\ref{COOLRATE}).  This reduces the problem to one
of evaluating two-dimensional integrals on $\bk$ for the appropriate
spin wave dispersion $\epsilon_\lambda(h,{\bf k})$.  

We note that in the special non-interacting case $\epsilon(h,{\bf k})
= h$, these expressions reduce to those for an ideal quantum
paramagnet, with the associated magnetocaloric effect:
\begin{eqnarray}
\Gamma^0_{\text{mc}}=\left(\frac{\partial T}{\partial H}\right)_S &=&
\frac{T}{H}
\end{eqnarray} 
Incidentally this is a general property of any system for which the
partition function $Z$ depends only on $H/T$, i.\,e.{} $F = -k_{\text
B}T \ln Z(H/T)$.

\subsubsection{Spin wave dispersion in the ferromagnet/saturated
paramagnet}
\label{sec:LSWdisp}

Since spin wave theory {\em assumes\/} broken spin rotation symmetry,
we can treat the spontaneously polarized FM phase for $h=0$ and the
saturated paramagnet for $h>h_{\text c}(\phi)$ on an equal footing.
Expanding about the maximally polarized state to ${\cal O}(1/S)$ we
find
\begin{eqnarray}
{\cal H} &=& NE_0 + \sum_{\bf k} \epsilon_{\rm
FM}(h,{\bf k}) a^\dagger_{\bf k} a^{\phantom\dagger}_{\bf k} +
{\cal O}(1/S^2)
\end{eqnarray}
where 
\begin{eqnarray}
E_0 = 2 (J_1 + J_2)S^2 - h S
\end{eqnarray}
is the classical ground state energy per spin, which is equal to
$U_{\text{mf}}(T=0)$ in Eq.~(\ref{UEN2}).  The spin wave dispersion has
a single branch given by:
\begin{eqnarray}
\epsilon_{\rm FM}(h,{\bf k}) = h -4J_1S[1 - \gamma({\bf k})] -4J_2S[1 -
\overline\gamma({\bf k})] 
\end{eqnarray}
where 
\begin{eqnarray}
    \gamma({\bf k}) &=& \frac{1}{2}( \cos k_x + \cos k_y)
    \quad\mbox{and}
    \\
    \overline\gamma({\bf k}) &=& \cos k_x \cos k_y 
\end{eqnarray}
We then have simply
\begin{eqnarray}
\frac{\partial\epsilon_{\rm FM}(h,{\bf k})}{\partial h} = 1
\end{eqnarray}
i.\,e.{} a rigid shift of the entire dispersion with change in magnetic
field.

The dispersion will have a single (parabolic) minimum at ${\bf k} =
(0,0)$ in the FM phase (i.\,e.{} for $J_1 <0$, $J_2 < |J_1|/2$).
However in the saturated paramagnetic state, above the critical field
$h_{\text c}(\phi,J_{\text c})$, the minimum of the dispersion will be
at ${\bf k} = (\pi,\pi)$ where there is a NAF ground state, and at
${\bf k} = (\pi,0)$ (and symmetry points) where there is a CAF ground
state.  At the classical critical point $J_1 = 2J_2 > 0$ separating
NAF from the CAF, there are line zeros around the zone boundary $k_x =
\pm \pi$ and $k_y = \pm \pi$.  At the classical critical point $-J_1 =
2J_2 > 0$ separating FM from the CAF, there are line zeros for $k_x =
0$ and $k_y = 0$.  Note that these line zeros connect the different
minima of the dispersions between which this state must interpolate.\\

\subsubsection{Spin wave dispersion in the canted NAF}

Expanding about a canted NAF with ordering vector $(\pi,\pi)$ and
canting angle $\theta_{\text c}$
we find
\begin{widetext}
    \begin{equation}
	\label{BILNAF}
	{\cal H} = NE_0 + \sum_{\bf k}\bigl[ 
	A(h,{\bf k}) \left( a^\dagger_{\bf k} a^{\phantom\dagger}_{\bf k} +
	b^\dagger_{\bf k} b^{\phantom\dagger}_{\bf k} \right)
	+ B(h,{\bf k}) \left( a^\dagger_{\bf k} b^{\dagger}_{\bf -k} + 
	a^{\phantom\dagger}_{\bf k} b^{\phantom\dagger}_{\bf -k} \right)
	+ C(h,{\bf k}) \left( a^\dagger_{\bf k} b^{\phantom\dagger}_{\bf k}
	+ b^\dagger_{\bf k} a^{\phantom\dagger}_{\bf k} \right)\bigr]
	+ {\cal O}(E_0/S^2)
    \end{equation}
\end{widetext}
The classical ground state energy per site is given by
\begin{eqnarray}
E_0 = 2J_1S^2 \cos \theta_{\text c} + 2J_2S^2 - h S\cos (\theta_{\text c}/2)
\end{eqnarray}
which is identical to $U_{\text{mf}}(T=0)$ of Eq.~(\ref{UEN2}).
Minimizing this energy fixes the canting angle (Eq.~(\ref{MFANGLE}))
\begin{eqnarray}
    \label{eqn:htheta}
\frac{\theta_{\text c}}{2} &=& \cos^{-1} \left(\frac{h}{8J_1S}\right)
\end{eqnarray}
\begin{figure*}
    \centering
    \hfill
    \includegraphics[height=0.30\textwidth]{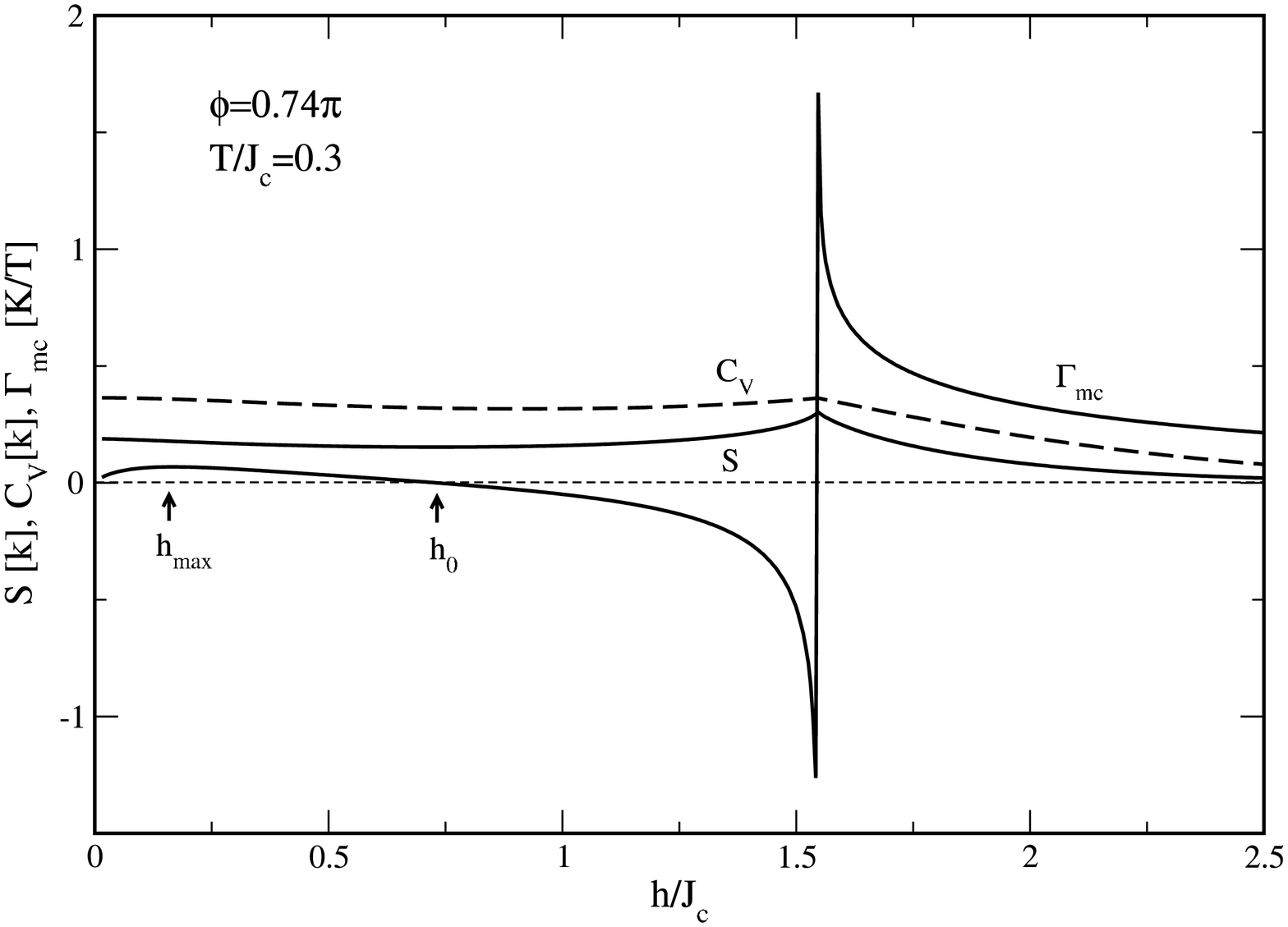}
    \hfill\includegraphics[height=0.30\textwidth]{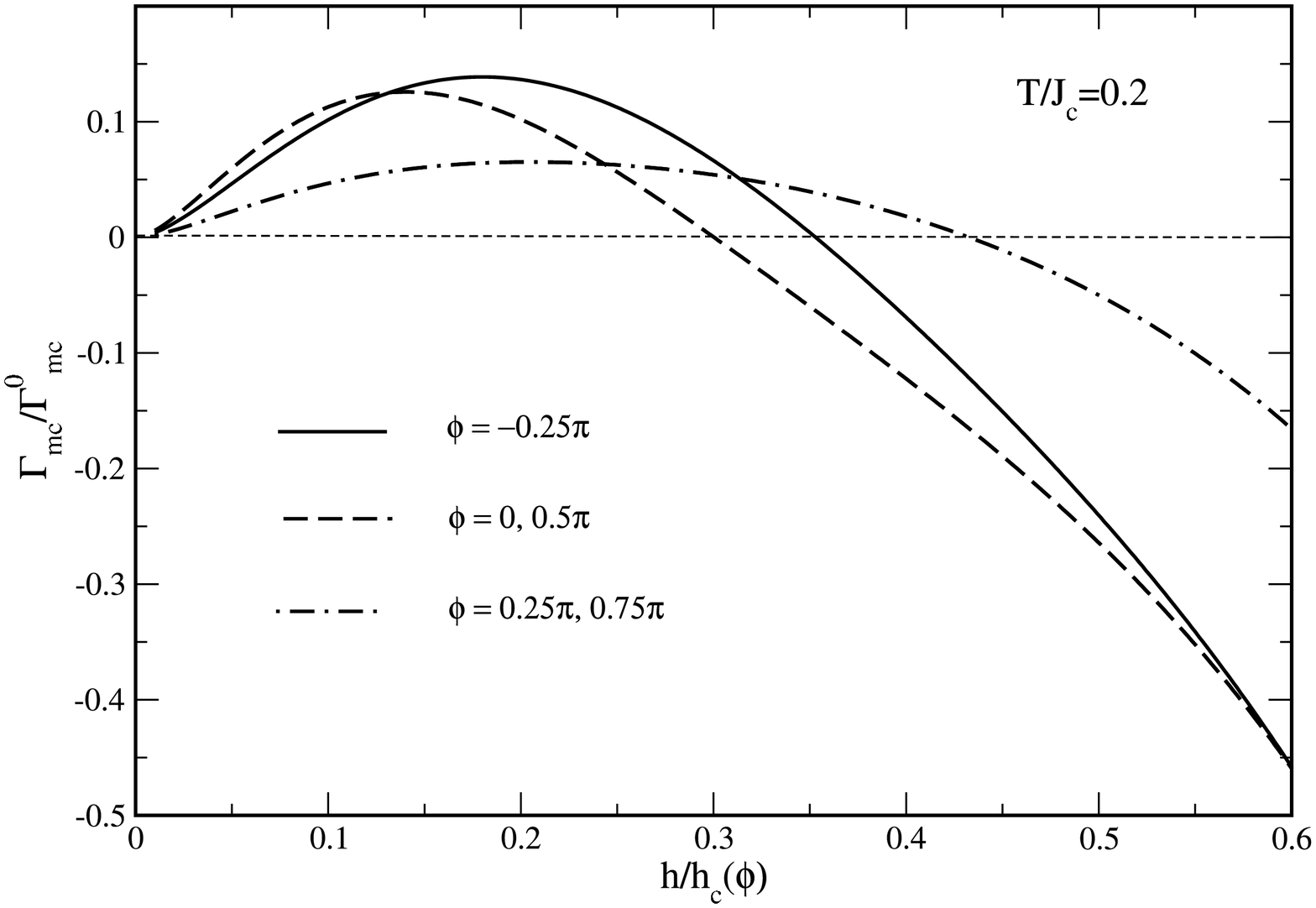}\hfill\null
    \caption{Left panel: Entropy, specific heat and MCE for $\phi$ in
    the CAF regime.  The MCE has a maximum at $h_{\text {max}}$ and
    then changes sign at $h_0$.  At the saturation field $h_{\text c}$
    another sign change associated with a spike-like singularity
    appears due to the field gradient of the entropy.  Right panel:
    Normalized MCE cooling rate
    $\Gamma_{\text{mc}}/\Gamma^0_{\text{mc}} =(H/T)\Gamma_{\text{mc}}$
    for five different frustration angles in the NAF ($\phi=-0.25\pi,
    0$) and CAF ($\phi=0.25\pi, 0.5\pi, 0.75\pi$) regime and shown for
    moderate fields.  The field is normalized to the saturation field
    $h_{\text c}(\phi)$ given in Fig.~\ref{fig:scaling} (right).  Low
    field maximum of the MCE at $h_{\text{max}}(\phi)$ and sign change
    at $h_0(\phi)$ are clearly seen to occur for all frustration
    angles.  Pairwise equalities of the MCE are observed due to the
    symmetries of the spinwave spectrum with respect to $\phi$.}
    \label{fig:swtherm}
\end{figure*}
where $(\theta_{\text c}/2)$ is measured relative to the magnetic
field direction z (i.\,e.{} the FM has $\theta_{\text c} \equiv 0$,
the NAF $\theta_{\text c} \equiv \pi/2$).  Exactly the same expression
follows from the requirement that the spin wave expansion contains no
terms linear in bosons. Eliminating the magnetic field through 
Eq.~(\ref{eqn:htheta}), we find:
\begin{eqnarray}
    A(h,{\bf k}) \pm C(h,{\bf k})&=& 4J_1S\left[ 1 \pm
    \cos^2(\theta_{\text c}/2)\gamma({\bf k})\right] 
    \nonumber\\&&{}
    - 4J_2S\left[ 1 -\overline\gamma({\bf k})\right]\\
    B(h,{\bf k}) &=& - 4J_1S \gamma({\bf k}) \sin^2(\theta_{\text c}/2)
    \nonumber
\end{eqnarray}
These expressions still depend on the applied magnetic field through
the canting angle $\theta_{\text c}(h)$.  The bilinear form of
Eq.~(\ref{BILNAF}) (Appendix B) may be diagonalized by a Bogoliubov
transformation to give
\begin{eqnarray}
    \label{HAMNAF}
    {\cal H} &=& NE_0 + NE_{\text{zp}}+ \sum_{\bf k, \lambda=\pm}
    \epsilon_\lambda (h,{\bf k})
    \alpha^{\dagger}_{\lambda \bf k} \alpha^{\phantom\dagger}_{\lambda
    \bf k}
    \nonumber\\&&{}
    + {\cal O}(1/S^2)
\end{eqnarray}
where 
\begin{eqnarray}
    \label{eqn:dispersion}
\epsilon_\pm (h,{\bf k}) &=& \sqrt{[A (h,{\bf k}) \pm C(h,{\bf k})]^2
- B(h, {\bf k})^2}
\label{DISPNAF}
\end{eqnarray}
The two-fold degeneracy of the spin wave dispersion of the NAF is
lifted by the applied magnetic field.  In the (physical) magnetic
Brillouin zone centered on ${\bf k} = (\pi, \pi)$, the Goldstone mode
is $\epsilon_+ (h,{\bf k})=0$, while $\epsilon_- (h,{\bf k})=h$ has a
finite gap.  Since spin waves are not eigenstates, there is now a
zero-point energy term in the energy E$_{\text{zp}}$.

In order to calculate the rate of change of magnetization with 
magnetic field in Eq.~(\ref{MGRADLSW}), we also need the field 
derivative of the dispersion.  
Considering explicitly $\epsilon_+(h,\bk)$, we obtain 
\begin{eqnarray}
    \label{eqn:peter}
\frac{\partial\epsilon_+(h,\bk)}{\partial
h}=\frac{1}{\epsilon^+_\bk}\frac{2C_\bk}{h}(A_\bk+C_\bk-B_\bk)
\end{eqnarray}
As far as the periodicity of dispersions is concerned, note that
translation by {\bf Q}=($\pi,\pi$) leads to
\begin{eqnarray}
    \label{eqn:bigQ}
    A(h,{\bf k} + {\bf Q}) + C(h,{\bf k} + {\bf Q}) &=& 
    A(h,{\bf k}) - C(h,{\bf k})\no\\
    B(h,{\bf k} + {\bf Q}) &=& - B(h,{\bf k}) 
\end{eqnarray}
therefore the two $\epsilon_\pm$ modes are simply interchanged by
translation through the NAF ordering vector {\bf Q}=$(\pi, \pi)$.
This means that in simple thermodynamic averages one can work with a
single mode in the full square lattice BZ---e.\,g.{}
$\epsilon_+(h,\bk)$---rather than with the two (physically distinct)
modes in the smaller magnetic BZ. The $\epsilon_+ (h,{\bf k})$ spin
wave dispersion for NAF in the full paramagnetic BZ is shown in
Fig.~\ref{fig:omega1NAFcanted} (left panel).
\begin{figure*}
    \centering
    \hfill
    \includegraphics[height=0.30\textwidth]{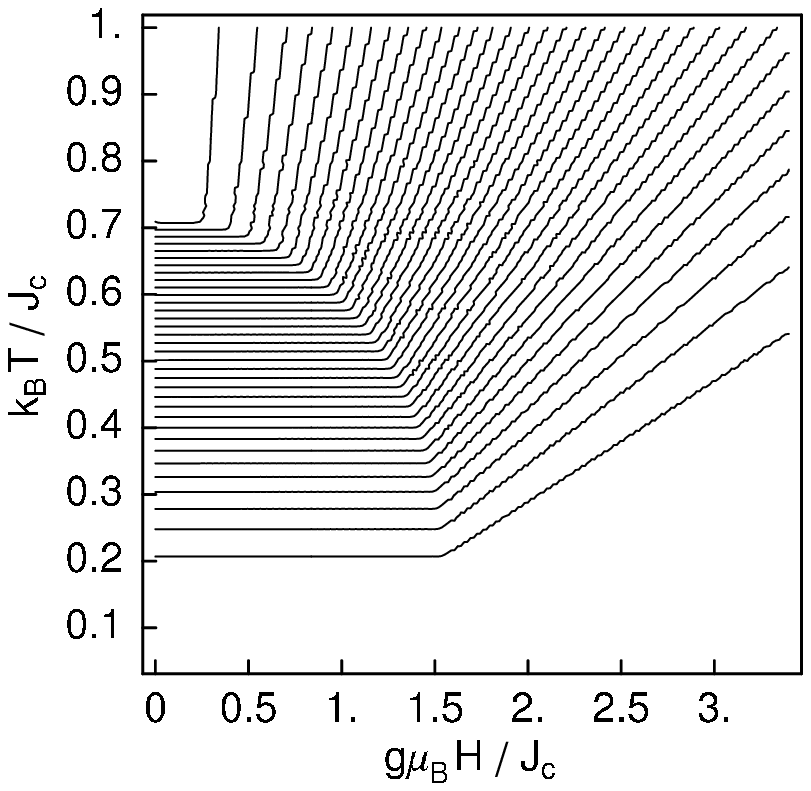}
    \hfill\includegraphics[height=0.30\textwidth]{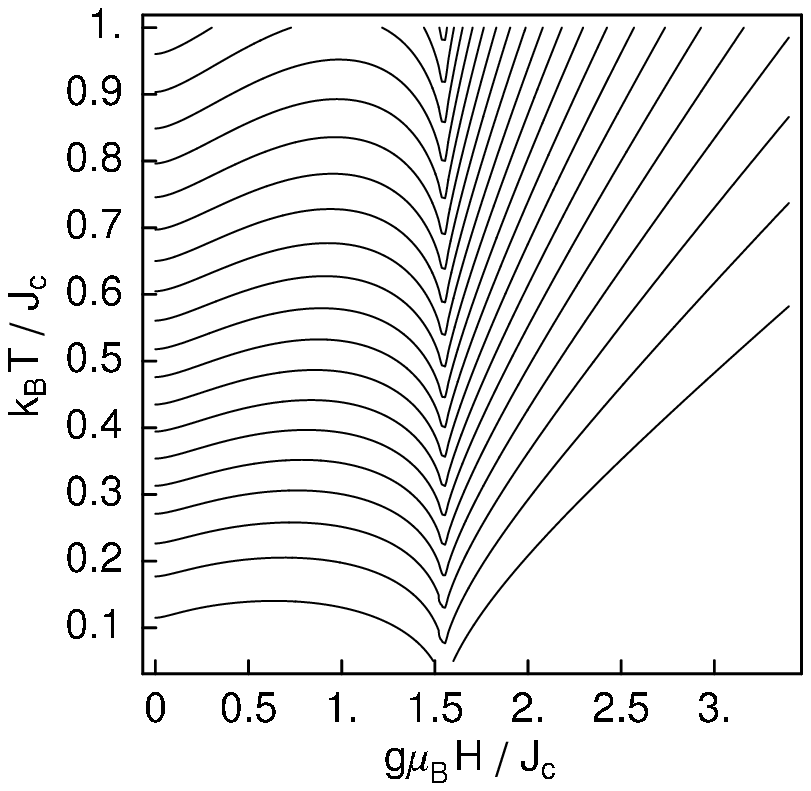}\hfill\null
    \caption{Contour plots of the entropy S(T,H) in mean field
    approximation (left panel) and spin wave approximation (right
    panel) in steps of $\Delta S=0.02\,N_{\text A}k_{\text B}$ and
    $0.05\,N_{\text A}k_{\text B}$ respectively.  The former shows a
    temperature dependent critical field $H_{\text c}$ but no
    structure of the entropy.  The latter has a constant critical
    field but exhibits the cusp structure around $H_{\text c}$ which
    is responsible for the sign change and peaks of the MCE in
    Fig.~\ref{fig:swtherm}.  These contour plots should be compared
    with the results from FTLM in Fig.~\ref{fig:isosandcv} (left
    panel) which qualitatively exhibit both features.}
    \label{fig:scont_ht}
\end{figure*}

\subsubsection{Spin wave dispersion in the canted CAF}

The ground state energy per spin is now given by 
\begin{eqnarray}
    E_0 &=& J_1S^2 [1 + \cos \theta_{\text c}]
    + 2J_2S^2 \cos\theta_{\text c}
    \nonumber\\&&{}
    - h S \cos(\theta_{\text c}/2)
\end{eqnarray}
which again is equal to $U_{\text{mf}}(T=0)$ in Eq.~(\ref{FUEN}).
Minimizing this energy leads to a canting angle (cf.{}
Eq.~\ref{MFANGLE})
\begin{eqnarray}
    \frac{\theta_{\text c}}{2} = \cos^{-1}\left(\frac{h}{4 J_1 S +8
    J_2 S}\right)
     \label{CANTCAF}
\end{eqnarray}
Once again we obtain spin waves with a dispersion of the form 
Eq.~(\ref{eqn:dispersion}).  
After elimination of the field using Eq.~(\ref{CANTCAF}), the coefficients 
of the spin wave expansion are given by
\begin{eqnarray}
\label{ABCCAF}
A(h,{\bf k}) &=&  {\phantom-}2 S [2 J_2 + J_1 \cos  k_y]\\
B(h,{\bf k}) &=& -2 S  [J_1 + 2 J_2 \cos  k_y] \cos k_x
\sin^2(\theta_{\text c}/2)\no\\
C(h,{\bf k}) &=&   {\phantom-}2 S  [J_1+2 J_2 \cos  k_y] \cos k_x
\cos ^2(\theta_{\text c}/2)\no 
\end{eqnarray}
These coefficients once again satisfy the relation
Eq.~(\ref{eqn:bigQ}) with {\bf Q}=$(\pi,0)$, i.\,e.{} the
$\epsilon_\pm$ modes are interchanged under translation through the
magnetic ordering vector.  Therefore simple averages can again be
calculated for a single mode in the full paramagnetic (square lattice)
BZ. As in the NAF case, the field gradient of spin wave energies is
given by Eq.~(\ref{eqn:peter}).
The $\epsilon_+ (h,{\bf k})$ spin wave dispersion for the CAF in the 
full paramagnetic BZ is shown in Fig.~\ref{fig:omega1NAFcanted} (right panel).

\section{Discussion of the analytical results and comparison with the
numerical findings}
\label{sect:DISCUSSION}

The typical field dependence of entropy, specific heat and MCE in the
linear spin wave approximation as calculated from
Eqs.~(\ref{COOLRATE}, \ref{SLSW}, \ref{MGRADLSW}, \ref{SPECLSW}) are
shown in Fig.~\ref{fig:swtherm}.  Entropy, specific heat and MCE are
all smooth functions of magnetic field {\em except\/} at the critical
field $h_{\text c}$ at which there is a (2nd-order) phase transition
between the paramagnet and canted N\'eel phases.

The most striking feature of these predictions is the double spike in
the MCE at $h_{\text c}$.  This is a generic feature of a 2nd order
phase transition between paramagnetic and ordered phases in applied
field~\cite{garst:05} and has previously been seen in Monte Carlo
simulations of the classical Heisenberg model~\cite{zhitomirsky:03}.
This sudden and sharp change in sign of the MCE can easily be
understood in terms of the contours of fixed entropy (adiabats),
discussed below.  It is accompanied by closely related cusps in the
entropy and heat capacity, peaked at $h_{\text c}$.

In general, as we would expect, entropy and specific heat have a much
weaker field dependence in the ordered phase below $h_{\text c}$ than
in the disordered phase above it.  The entropy of the (gapped)
paramagnetic phase falls rapidly in applied magnetic field, while the
N\'eel phase responds to magnetic field by canting, at nearly constant
entropy.  As a result the typical (absolute) value of the MCE is much
larger above $h_{\text c}$ than below.  These features of the LSW
predictions are reminiscent of the mean field theory, as illustrated
in Fig.~\ref{fig:MFTHERMH}.

For sufficiently large $h \gg T, J_{\text c}$ we must (and do) recover
the response of an isolated paramagnetic spin in either approximation.
Where the LSW predictions differ from those of mean field theory is in
the singular features at $h_{\text c}$, and in the presence of a
finite MCE in the ordered phase.  This is positive for $h \to 0$,
exhibits a shallow maximum at a characteristic field $h =
h_{\text{max}}$, changes sign at another characteristic field $h=h_0$,
before exhibiting the dramatic double spike at $h=h_{\text
c}$---irrespective of which ordered phase is considered.  The absolute
values of these critical fields, and the form of the anomalies at
$h_{\text c}$, {\em do\/} however depend on the structure of the low
energy spin spectrum, and therefore on frustration through $\phi$.

\begin{figure*}
    \includegraphics[width=0.50\textwidth]{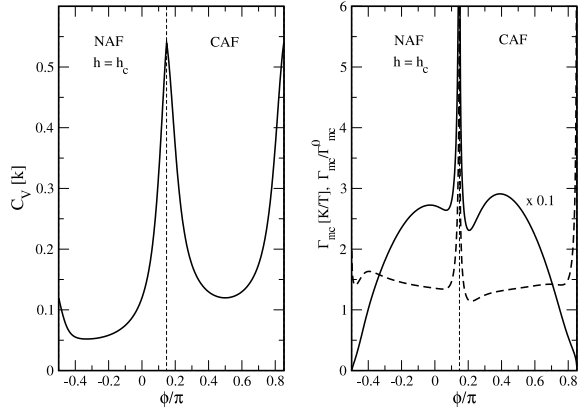}
    \caption{Thermodynamic quantities slightly above the critical
    field $h_{\text c}(\phi)$ as function of $\phi$ for temperature
    $k_{\text B}T/J_{\text c}=0.2$.  Here we used $\Delta_h=h-h_{\text
    c}=10^{-3}h_{\text c}$.  Left: Specific heat shows enhancement at
    the NAF/CAF boundary.  Right: Bare (dashed line) and normalized
    (solid line) magnetocaloric effect as function of $\phi$.
    Giant peak at the NAF/CAF boundary occurs due to gapless
    LSW modes along lines in the BZ. The overall behavior of the
    normalized $\Gamma_{\text{mc}}/\Gamma^0_{\text{mc}}=(H_{\text
    c}(\phi)/T)\Gamma_{\text{mc}}$ follows the $\phi$-dependence of
    the critical field (Fig.~\ref{fig:scaling}(right)).}
    \label{fig:swcoolphi}
\end{figure*}
The difference between mean field and spin wave results becomes most
obvious in a comparison contour plot of the entropy $S(T,H)$ shown in
Fig.~\ref{fig:scont_ht}.  The former (left panel) has a temperature
dependent critical field which vanishes at the (mean field) transition
temperature.  The entropy does not have any structure below $h_{\text
c}$ due to the field independent molecular field splitting $\Delta$ in
Eq.(~\ref{ROTATE}).  In spin wave approximation (right panel) the
critical field is temperature independent, but the entropy contours
show a typical cusp structure around $h_{\text c}$ with a maximum at
$h_0$ further down which is caused by the excitation of low energy
spin waves (cf.{} Ref.~\onlinecite{kalva:76,kalva:78}).  According to
the definition of $\Gamma_{\text{mc}}$ in Eq.~(\ref{COOLRATE}) this
immediately translates into the sign change of $\Gamma_{\text{mc}}$ at
$h_0$ and its enhancement around $h_{\text c}$.

The behavior of the (normalized) MCE in the low to moderate field
regime is presented on an enlarged scale in Fig.~\ref{fig:swtherm}
(right panel) for typical values of $\phi$ in the NAF/CAF region.  We
notice a considerable variation of the characteristic fields
$h_{\text{max}}(\phi)$ and $h_0(\phi)$ with the frustration angle.
The maximum MCE, $\Gamma_{\text{mc}}(h_{\text{max}})$ for $k_{\text
B}T/J_{\text c} = 0.2$ is of the order of ten percent of the
paramagnetic value $\Gamma^0_{\text{mc}}(h_{\text{max}})$.
Furthermore a symmetry in the $\phi$ dependence is obvious: Firstly
the MCE is invariant under reflection at the axis $\phi=0.5\pi$ or
$J_1=0$ when both values of $\phi$ lie in the CAF sector.  This is
obvious from the spin wave dispersion Eq.~(\ref{DISPNAF},\ref{ABCCAF})
in the CAF regime which is invariant under the simultaneous
transformation $(J_1,J_2)\rightarrow (-J_1,J_2)$ and
$(k_x,k_y)\rightarrow(k_x+\pi,k_y+\pi)$.  Since the MCE is obtained by
integration over the whole BZ, $\Gamma_{\text{mc}}$ is unchanged under
sign reversal $J_1\rightarrow -J_1$.  Secondly the MCE for $\phi=0$
and $\phi=0.5\pi$ are equal, i.\,e.{}, it is invariant under the
replacement $(J_1,0)\rightarrow(0,J_2$).

Finite size effects prevent the characteristic fields
$h_{\text{max}}(\phi)$ and $h_0(\phi)$ from being identified in FTLM
calculations, as shown in Fig.~\ref{fig:comparison}.  Nevertheless a
negative MCE at moderate fields is clearly compatible with the
numerical results.  In practice, for a finite size cluster, each of
the ground state level crossings shown in
Fig.~\ref{fig:levelcrossings} shows up as a separate ``phase
transition'' in the FTLM results for the MCE, with associated positive
and negative spikes in Fig.~\ref{fig:comparison}.  For this reason,
the sign of the MCE remains ambiguous.  However for the saturated
paramagnetic phase, where there is no further level crossing in the
ground state, the sign of the MCE is correctly resolved and a
pronounced enhancement is seen approaching $h_{\text c}(\phi)$ from
above.

This is in good qualitative agreement with the predictions of LSW
theory, where it is clear from Fig.~\ref{fig:swtherm} that the largest
positive MCE is to be expected {\em just above\/}~$h_{\text c}(\phi)$.
This maximum arises from the closing of the spin wave gap in the fully
polarized phase when $h\rightarrow h_{\text c}^+$.  It occurs for
fields slightly above $h_{\text c}(\phi)$ because temperature acts to
``round'' the sharp cusp in the entropy contours at the critical field
--- c.f. Fig.~\ref{fig:scont_ht}.  Needless to say, the low energy
excitations responsible for the sharply diverging peaks seen in LSW
results are not accurately described by a cluster of 24 sites, and so
it makes little sense to compare FTLM and LSW predictions at a
quantitative level.

While the structure of the MCE is generic to a (second order) phase
transition, the details depend strongly on the amount of frustration
present.  If the frustration angle is deeply within one of the ordered
sectors, the softening of the spin waves at $h\rightarrow h_{\text
c}^+$ occurs at the wave vector of the low field AF structure
$(\pi,\pi)$ or $(0,\pi)$.  However if the frustration angle approaches
the transition regions CAF/NAF and CAF/FM, the softening will occur
along the whole line in the BZ connecting the wave vectors of the
competing structures.  This is reminiscent of, but less dramatic than,
the situation in certain geometrically frustrated magnets where the
spin gap closes simultaneously at $h_{\text c}$ for an entire branch
of excitations across the BZ, leading to the condensation of a
macroscopic number of localized magnon modes~\cite{zhitomirsky:03}.
In the present case one should therefore expect a strong enhancement
of $\Gamma_{\text{mc}}(h\rightarrow h_{\text c}^+)$ for $\phi$ close
to one of the above boundary regions.

The same holds for the specific heat.  In Fig.~\ref{fig:swcoolphi}
(left panel) we show the peak value $C_V(h^+_{\text c},\phi)$ as
function of frustration angle.  Indeed the specific heat shows a
strong enhancement for $\phi\simeq 0.15\pi$ (CAF/NAF) and $\phi \simeq
0.85\pi$ (CAF/FM) of considerable width in $\phi$.  This is in good
qualitative agreement with the FTLM results for finite clusters
presented in Fig.~\ref{fig:mceandcvsat} (left panel).  The anomaly at
the FM/NAF boundary on the other hand is much smaller.

On the right panel of  Fig.~\ref{fig:swcoolphi} the corresponding plot 
for the MCE is shown.  The dashed line shows the bare MCE coefficient
$\Gamma_{\text{mc}}(h\rightarrow h_{\text c}^+)$ as function of
$\phi$.  It is almost constant except at the phase boundaries where
again a large, but much narrower peak appears.  This is not
immediately obvious since $C_V$ enters in the denominator of the
expression for $\Gamma_{\text{mc}}$ in Eq.~(\ref{COOLRATE}).  In fact
on approaching $\phi/\pi\simeq0.15$ the MCE slightly decreases, only
very close to the value when the spin wave dispersion softens along
the line $(\pi,0)$--$(\pi,\pi)$ and equivalent ones in the BZ a very
sharp spike appears.  This is due to the fact that the spin wave
softening leads to a stronger increase of the magnetization gradient
(Eq.~(\ref{MGRADLSW})) as compared to the increase of $C_V$ from
Eq.~(\ref{SPECLSW}).  The full line in Fig.~\ref{fig:swcoolphi} (right
panel) shows the normalized MCE. Aside from the phase boundaries where
again sharp peaks appear it is largely determined by the behavior of
the saturation field (Fig.~\ref{fig:scaling}) since the bare
$\Gamma_{\text{mc}}(h\rightarrow h_{\text c}^+)$ is roughly constant
in $\phi$.

Deep within the ordered phases the degree of enhancement of the MCE
relative to an ideal paramagnet is chiefly controlled by the
saturation field $h_{\text c}(\phi)$.  The FTLM and LSW predictions
are therefore in excellent qualitative agreement (c.f.
Fig.~\ref{fig:mceandcvsat} and Fig.~\ref{fig:swcoolphi}).  However,
once again, finite size effects prevent the FTLM method from capturing
the full extent of the anomalous enhancement of the MCE in the highly
frustrated regions at the borders of the CAF phase.  These may in
practice be overestimated by LSW theory, since it takes no account of
new non-magnetic phases stabilized by fluctuations.
\begin{figure}
    \includegraphics[width=0.35\textwidth]{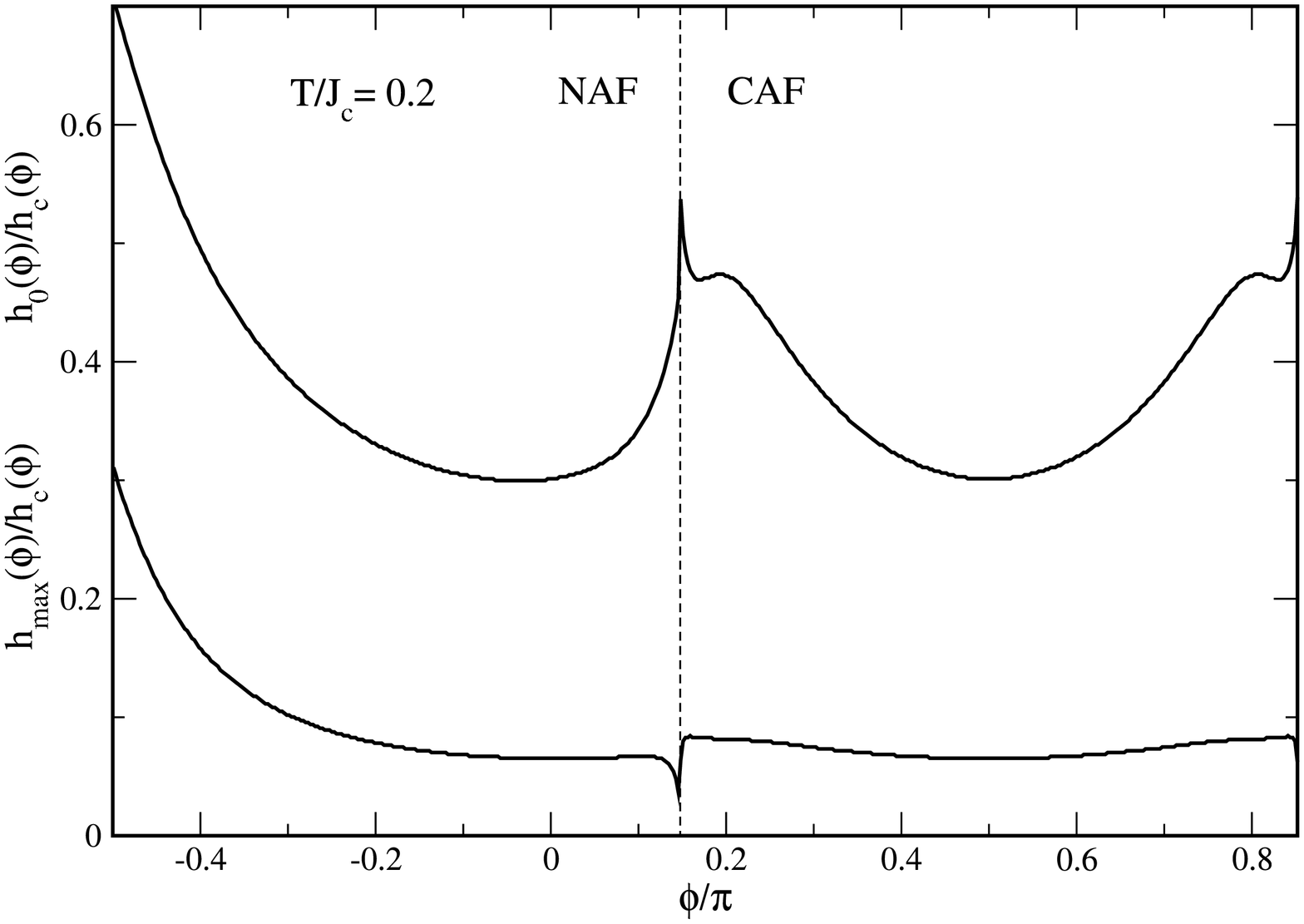}
    \caption{In this figure the subcritical fields $h_{\text{max}}(\phi )$
    and $h_0(\phi)$ where $\Gamma_{\text{mc}}(\phi)$ in
    Fig.~\ref{fig:swtherm} (right panel) is maximal or changes its
    sign respectively are plotted as function of the frustration
    angle.  In the CAF phase the characteristic fields are symmetric
    with respect to $\phi=0.5\pi$.}
    \label{fig:swcoolhcrit}
\end{figure}
None the less we can gain further insight into the strong enhancement of
$\Gamma_{\text{mc}}(h\rightarrow h_{\text c})$ and the singular
peak at $\phi_{\text c}$ in Fig.~\ref{fig:swcoolphi} within
the LSW approach by expanding the spin wave energies of the fully
polarized phase around the incipient ordering vector.  Explicitly, for
$\phi$ in the classical NAF sector we have
\begin{equation}
\epsilon_\bq=\Delta_h+a_2(q^2_x+q^2_y)-a_4(q^4_x+q^4_y)+a'_4q_x^2q_y^2
\label{DISPEXP}
\end{equation}
where $\bq=\bk-\bQ$ is the distance from the NAF vector
$\bQ=(\pi,\pi)$ and $\Delta_h=h-h_{\text c}$ is the excitation gap
with the NAF critical field $h_{\text c}=8SJ_1$, The expansion
coefficients are given by $a_2=2a_4=S(J_1-2J_2)$ and $a'_4=SJ_2$.
This expansion may be inserted into Eqs.~(\ref{MGRADLSW},
\ref{SPECLSW}) and the integration performed approximately
analytically.  It is assumed that only modes with an energy
$\epsilon_\bq < k_{\text B}T$ contribute appreciably to the integral.
In performing the integration one has to distinguish two cases: If one
is within the NAF sector the second order coefficient $a_2$ is
nonzero.  If one is at the boundary to the CAF regime $a_2=0$ and the
dispersion is determined by the mixed fourth oder coefficient $a'_4$.

Performing the integration in this (classical) limit \mbox{$k_{\text
B}T \gg \Delta_h$} one obtains the approximate expressions
\begin{eqnarray}
    \frac{\Gamma_{\text{mc}}}{\Gamma^0_{\text{mc}}}&\simeq&
    \left(\frac{h_{\text c}}{k_{\text B}T}\right)
    \ln\left(\frac{k_{\text B}T}{\Delta_h}\right)
    \\&&{}
    \mbox{for}\quad 2J_2 < J_1\quad \mbox{(NAF)}\label{classical1}
    \nonumber\\
    \frac{\Gamma_{\text{mc}}}{\Gamma^0_{\text{mc}}}&\simeq&
    2\left(\frac{h_{\text c}}{k_{\text B}T}\right)
    \left(\frac{SJ_2}{\Delta_h}\right)^\frac{1}{2}
    \ln\left(\frac{k_{\text B}T}{(SJ_2\Delta_h)^\frac{1}{2}}\right)\label{classical2}
    \\&&{}
    \mbox{for}\quad 2J_2=J_1 \quad\mbox{(NAF/CAF)}
    \nonumber
\label{MCESCALE}
\end{eqnarray}
In the corresponding quantum limit $k_{\text B}T \ll \Delta_h$, both
the heat capacity and the rate of change of magnetization with
temperature have an activated behavior.  However this cancels in
between the numerator and denominator of Eq.~(\ref{COOLRATE}) to give
\begin{eqnarray}
   \Gamma_{\text{mc}}
       &=&
       \frac{k_{\text B}T}{h-h_c}
\end{eqnarray}
regardless of the degree of frustration present in the model.  We note
that this is exactly the form predicted at a quantum critical point on
the basis of scaling arguments~\cite{garst:05}.

Returning to Eq.~(\ref{classical1},\ref{classical2})---in the first
case, inside the NAF sector, the divergence in the MCE for
$h\rightarrow h_{\text c}^+$ is only of a weak logarithmic type.
However at the classical boundary to the CAF sector (which widens into
the disordered regime due to quantum fluctuations) the singularity
becomes a much stronger one essentially of inverse square root type.
This is the reason that the MCE at $h=h^+_{\text c}$ shows a large
anomalous peak as function of $\phi$ when crossing the NAF/CAF
boundary.  Because of the vanishing of the second order term in
Eq.~(\ref{DISPEXP}) the dispersion has a saddle point at $\bf Q$
leading to a large DOS of low energy spin waves for $2J_2\simeq J_1$
and therefore a stronger algebraic divergence of the MCE at $h_{\text
c}$ appears.  The same arguments hold for the CAF/FM boundary.

However a word of caution is appropriate.  In our spin wave
calculations we assumed that the classical magnetic phases are stable
throughout the phase diagram.  Strictly speaking this is not true.  As
indicated in Fig.~\ref{fig:OSphase} in the shaded sectors around the
CAF/NAF and CAF/FM phase boundary quantum fluctuations lead to
instability of magnetic order and select a different nonmagnetic order
parameter, presumably staggered dimer~\cite{sushkov:01} and
spin-nematic~\cite{shannon:06}.  While the broad features of our
theory can be trusted, a truly quantitative theory for the nonmagnetic
sectors would require to start from the proper order parameter and
their associated elementary excitations.  In real materials,
sufficiently close to $h_{\text c}(T)$, the critical behavior of the
MCE will also be sensitive to the details of interlayer coupling and
magnetic anisotropy.  Both refinements remain as an outstanding
challenge.

It is also instructive to track the low to moderate field anomalies in
$\Gamma_{\text{mc}}(h,\phi)$ as function of $\phi$ around the phase
diagram.  In Fig.~\ref{fig:swcoolhcrit} the maximum field
$h_{\text{max}}(\phi)$ and the field $h_0(\phi)$ at which the MCE
changes sign are plotted as function of frustration angle, normalized
to the saturation field $h_{\text c}(\phi)$.  While
$h_{\text{max}}(\phi)/ h_{\text c}(\phi)$ is rather constant
throughout most of the range of angles, $h_{0}(\phi)$/$h_{\text
c}(\phi)$ shows considerable variation in $\phi$.  The two
characteristic fields are again symmetric around $\phi$=0.5$\pi$ or
$J_1$=0 for the same reasons as explained above.  Note that the
overall double-minimum structure of $h_{0}(\phi)$/ $h_{\text c}(\phi)$
as function of $\phi$ prevents its use as a criteria to resolve the
ambiguity of frustration angles that appears in zero-field
thermodynamic considerations mentioned in Sec.~\ref{sect:J1J2MODEL}.
It is obvious from the FTLM results in Fig.~\ref{fig:comparison}
(right panel), that the temperature dependence of the (normalized) MCE
above $h_{\text c}$ is strongly suppressed as $h$ increases.  This
effect can also be understood from the LSW calculations.
Approximating Eqs.~(\ref{MGRADLSW},\ref{SPECLSW}) for small and large
temperatures we obtain the ratio of the low and high temperature
(normalized) MCE as function of $h>h_{\text c}$:
\begin{eqnarray}
    \frac{\hat{\Gamma}_{\text{mc}}(T\ll J_{\text c})}%
    {\hat{\Gamma}_{\text{mc}}(T\gg J_{\text c})} \simeq
    \frac{\sum_\bk\epsilon(h,\bk)}
    {\sum_\bk\epsilon^2(h,\bk)\sum_\bk\epsilon^{-1}(h,\bk)}
\label{GRATIO}
\end{eqnarray}
As long as the field is not too far above $h_{\text c}$ there is still
a considerable dispersion in $\epsilon(h,\bk)$ and the above ratio is
larger than one (Fig.~\ref{fig:comparison}), i.\,e.{}
$\hat{\Gamma}_{\text{mc}}$ is $T$-dependent.  Once $h\gg h_{\text c}$
however the dispersion is negligible compared to the gap energy
$\Delta_h$ and the ratio in Eq.~(\ref{GRATIO}) approaches one,
i.\,e.{} we recover the behavior of an ideal paramagnet.

\begin{figure}
    \includegraphics[width=0.35\textwidth]{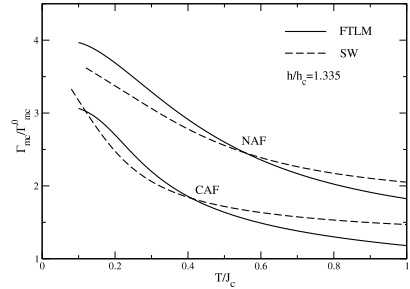}
    \caption{Normalized MCE as function of temperature for
    above-critical field calculated with FTLM an spin wave (LSW)
    method.  Frustration angles are $\phi=0$ (NAF) and $\phi/\pi=0.74$
    (CAF) corresponding to the Sr compound.  For $T\gg J_{\text c}$
    the FTLM result converges rapidly to 1.  The LSW result for finite
    $h/h_{\text c}$ has an asymptotic large temperature value
    different form one, only in the limit $h\gg h_{\text c}$ it also
    approaches one.}
    \label{fig:MCETNAFCAF}
\end{figure}

A comparison between FTLM and LSW predictions of the temperature
dependence of the MCE for fields safely above $h_{\text c}$ (in order
to avoid the logarithmic singularity at $h_{\text c}$) is given in
Fig.~\ref{fig:MCETNAFCAF}.  There is a reasonable agreement in both
magnitude and qualitative $T$-dependence.  Note however that the LSW
approximation becomes unreliable when $T$ approaches $J_{\text
c}$/$k_{\text B}$ and too many spin wave modes are thermally excited.

\section{Summary and Conclusion}

\label{sect:CONC}

We have investigated the magnetocaloric properties of the $J_1$-$J_2$
model using the FTLM method for finite clusters and spin wave analysis
starting from the classical magnetic structures.  The one-magnon
critical field or saturation field obtained from FTLM agrees well with
the spin wave result.  Finite size scaling results suggest that close
to the CAF/FM boundary, the true critical field is determined by a
two-magnon instability.  This is consistent with the proposed
existence of a spin nematic ground state in this parameter
range~\cite{shannon:06}.

Both FTLM and spin wave results predict a strong enhancement in the
low temperature specific heat at the saturation field when the
frustration angle crosses the phase boundaries.  This may be explained
by the large degeneracy of low lying states in these regions.  At a
constant intermediate field the specific heat exhibits a double peak
structure around the NAF/CAF boundary.  The entropy and specific heat
show only moderate field dependence below the saturation field
$h_{\text c}$.  This feature may already be understood in a mean field
approach where the entropy is strictly constant for all $h < h_{\text
c}$.

Likewise the strong enhancement of the MCE just above the saturation
field was investigated.  In the FTLM results, the MCE was enhanced by
up to a factor ten relative to an ideal paramagnetic (at temperatures
small compared to the energy scale $J_{\text c}$).  Surprisingly, the
largest enhancement (from FTLM, relative to an ideal paramagnet) does
not occur at the CAF/NAF boundary, but deep within the magnetically
ordered sectors.  This can be understood in terms of the anomalous
enhancement of the specific heat in the frustrated regions, which
enters into the denominator of the MCE (c.f. Eq.~(\ref{COOLRATE})).

The overall $\phi$ dependence of the MCE enhancement ratio reflects
that of the saturation field.  This is also true for the spin wave
results.  There, in addition, the enhancement is sharply peaked on the
CAF/NAF boundary appears.  This is due to the change of the
field-scaling behavior above the critical field from logarithmic to
inverse square-root when the boundary is crossed.  This feature is due
to the appearance of Goldstone modes along a line in the BZ when
$\phi$ has it critical value.  The MCE enhancement may also be
directly seen from the isentropics or adiabatic temperature curves
which exhibit a large slope above the saturation field.

Below the saturation field the results of the FTLM is strongly
constrained by finite size effects, however the spin wave analysis
provides considerable insight into the systematic properties of the
MCE. In absolute terms the MCE is very small---in fact much smaller
than the paramagnetic value.  This is a direct consequence of magnetic
order, and can easily be understood from the mean field picture of a
canted N\'eel state, for which entropy is constant as a function of
field.  The spin wave theory, however, predicts a flat maximum in the
MCE at low fields, followed by a sign change for subcritical fields.
These features are present throughout the phase diagram and the
characteristic fields are moderately $\phi$-dependent with a
double-minimum structure.  For this reason it is not likely that these
low-field MCE features are useful in the determination of the
frustration ratio.

Considerably above the critical field the temperature dependence
calculated from FTLM and spin wave theory show reasonable agreement.
At temperatures of the order of $J_{\text c}/k_{\text B}$ the
enhancement of the MCE is substantially reduced and the behavior of
the system crosses over to that of an ideal paramagnet.

The most pronounced discrepancies between FTLM and spin wave analysis
appear at the classical phase boundaries.  As explained above this may
be well understood in terms of the absence or presence of low lying
collective modes.  While the former method underestimates the MCE
anomalies at the phase boundaries, the latter overestimates them---in
fact it predicts a singular behavior.  A more advanced analytical
treatment would have to take into account the quantum nature of the
ground state around these boundaries, i.\,e.{} stacked dimer ($J_2>0$)
or spin nematic ($J_2<0$) and the proper associated excitation
spectrum.

The present analysis provides some interesting predictions for the
experimental investigation of the class of layered perovskites
discussed in the introduction.  Specifically we give detailed values
for the possible saturation fields which should be easily accessible
experimentally for SiO$_4$ and PO$_4$ vanadates.  According to
Fig.~\ref{fig:expfields}, these critical fields can be used to resolve
the ambiguity in parameterizing the model from its low-field
susceptibility and heat capacity, far more cheaply than, e.g. neutron
scattering.  Furthermore we predict a genuine sign change in the MCE
for subcritical fields which should be accessible to experiment.

So far as practical applications---for example in cryogen free
cooling---are concerned, the goal is to achieve as large a
magnetocaloric effect as possible, at as low a field as possible.
Here compounds not too far from the phase boundaries FM/NAF and CAF/FM
are the most promising because they combine a significant MCE
enhancement with very moderate saturation fields.  A detailed
treatment of entropy as a function of magnetic field in the nematic
phase occurring on the CAF/FM border remains an open challenge.
However the high density of low energy excitations and low saturation
field of this phase means that it looks {\em a priori\/} very
promising for magnetothermal applications.

\begin{acknowledgments}
    The authors are pleased to thank Tsutomu Momoi, Philippe
    Sindzingre and Matthias Vojta for helpful discussions.  N.S.
    acknowledges support under EPSRC Grant No.  EP/C539974/1.  P.T.
    acknowledges support by Deutsche Forschungsgemeinschaft under SFB
    463.
\end{acknowledgments}

\appendix

\section{Moment derivatives}
\label{sect:Appendix A} 

Here we give the explicit expressions of the moment derivatives
$(\partial \la S_i\ra/\partial \beta)$ ($i=\pa,\pe$) which are needed
for the mean field calculation of the magnetocaloric coefficient:
\begin{eqnarray}
    \label{STGRAD}
    \la S_\pe\ra'\equiv\frac{\partial\la S_\pe\ra}{\partial
    \beta}
    &=&[(1-A_{zz})B_x+A_{xz}B_z]/D\\
    \la S_\pa\ra'\equiv\frac{\partial\la S_\pa\ra}{\partial
    \beta}
    &=&[(1-A_{xx})B_z+A_{zx}B_x]/D
\end{eqnarray}
Where the determinant D is defined by
$D=[(1-A_{xx})(1-A_{zz})-A_{xz}A_{zx}]$ and  the coefficients $A_{ij}$
($i,j= x$ or $\pe$, $z$ or $\pa$) and $B_i$  are given by
\begin{eqnarray}
    \label{GRADCO1}
    B_x&=&
    \frac{\frac{1}{4}\hh_x}{\cosh^2\frac{1}{2}\beta\Delta}\\
    A_{xx}&=&
    \frac{a_\pe}{2\Delta}
    \frac{\beta\hh_\pe^2}{\Delta\cosh^2\frac{1}{2}\beta\Delta}
    +\frac{a_\pe}{\Delta}
    \left(\frac{\hh_\pa}{\Delta}\right)^2\tanh\frac{1}{2}\beta\Delta
    \no\\
    A_{xz}&=&
    -\frac{a_\pe}{2\Delta}
    \frac{\beta\hh_\pe\hh_\pa}{\Delta\cosh^2\frac{1}{2}\beta\Delta}
    +\frac{a_\pa}{\Delta}
    \frac{\hh_\pe\hh_\pa}{\Delta^2}\tanh\frac{1}{2}\beta\Delta
    \no
\end{eqnarray}
and likewise 
\begin{eqnarray}
    \label{GRADCO2}
    B_z&=&
    \frac{\frac{1}{4}\hh_z}{\cosh^2\frac{1}{2}\beta\Delta}\\
    A_{zz}&=&
    -\frac{a_\pa}{2\Delta}
    \frac{\beta\hh_\pa^2}{\Delta\cosh^2\frac{1}{2}\beta\Delta}
    -\frac{a_\pa}{\Delta}
    \left(\frac{\hh_\pe}{\Delta}\right)^2\tanh\frac{1}{2}\beta\Delta
    \no\\
    A_{zx}&=&
    \frac{a_\pe}{2\Delta}
    \frac{\beta\hh_\pe\hh_\pa}{\Delta\cosh^2\frac{1}{2}\beta\Delta}
    -\frac{a_\pe}{\Delta}
    \frac{\hh_\pe\hh_\pa}{\Delta^2}\tanh\frac{1}{2}\beta\Delta
    \nonumber
\end{eqnarray}
where $a_\pa, a_\pe$ are given in Eqn.~(\ref{ACOEFF}) and $\Delta,
\hh_\pa, \hh_\pe$ are defined in Eq.~(\ref{MOLFIELD}).

\section{Matrix form of the Hamiltonian}
\label{sect:Appendix B} 

In this appendix we give the explicit matrix form of the Hamiltonian
in Eq.~(\ref{BILNAF}) for the NAF structure which is bilinear in
bosonic  spin fluctuation operators.  It can be written as
\begin{widetext}
    \begin{eqnarray}
	{\cal H}^{\rm (2)} &=& \frac{1}{2} \sum_{\bf k} 
	\left(a^{\dagger}_{\bf k}, b^{\phantom\dagger}_{\bf -k},
	b^{\dagger}_{\bf k} ,a^{\phantom\dagger}_{\bf -k} \right)
	\left[ 
	\begin{array}{cc|cc}
	    A(h,{\bf k}) & B(h,{\bf k}) & C(h,{\bf k}) & 0\\
	    B(h,{\bf k}) &A(h,{\bf k}) & 0& C(h,{\bf k})\\ \hline
	    C(h,{\bf k}) & 0 &A(h,{\bf k}) & B(h,{\bf k}) \\
	    0 &  C(h,{\bf k}) & B(h,{\bf k}) &A(h,{\bf k}) 
	\end{array}
	\right]
	\left( 
	\begin{array}{l}
	    a^{\phantom\dagger}_{\bf k} \\
	    b^{\dagger}_{\bf -k} \\
	    b^{\phantom\dagger}_{\bf k}\\
	    a^{\dagger}_{\bf -k}
	\end{array}
	\right)
	-\sum_{\bf k}A(h,{\bf k})
    \end{eqnarray}
\end{widetext}
where the constant term (arising from spin commutation relations)
ensures that the zero point energy is negative.  Using a simple
coordinate rotation, we can reduce this matrix to a block diagonal form
with two $2\times2$ diagonal blocks given by
\begin{eqnarray}
  \left[ 
  \begin{array}{cc}
 A(h,{\bf k}) \pm C(h,{\bf k}) & B(h,{\bf k}) \\
  B(h,{\bf k})  & A(h,{\bf k}) \pm C(h,{\bf k})
  \end{array}
  \right] 
\end{eqnarray}
We can then solve each of the blocks using a separate, standard,
Bogoliubov transformation to obtain the diagonalized Hamiltonian in
terms of NAF spin wave modes as used in Eq.~(\ref{HAMNAF}).

Explicitly the complete transformation is given by
\begin{equation}
    \alpha_{\lambda\bf k}=\frac{u_{\lambda\bf k}}{\sqrt{2}}
    \left(b_{\bf k}+\lambda a_{\bf k}\right)+
    \frac{v_{\lambda\bf k}}{\sqrt{2}}
    \left(a^\dagger_{-\bf k}+\lambda b^\dagger_{-\bf k}\right)
\end{equation}
where $\lambda =\pm1$ is the branch index.  The coefficients
$u_{\lambda\bk}$ and $v_{\lambda\bk}$ of the transformation may be
obtained by direct insertion and the requirement that the off-diagonal
bilinear terms in the transformed operators $\alpha_{\lambda\bf k}$
vanish identically.  Alternatively the equations of motion for the
$\alpha_{\lambda\bf k}$ may be set up and required to describe free
motion with spin wave energy $\epsilon_{\lambda {\bf k}}$.  Both
methods lead to the same condition on the coefficients given by
\begin{eqnarray}
    2u_{\lambda\bf k}v_{\lambda\bf k}
    \left(A_{\bf k}+\lambda C_{\bf k}\right) =
    \left(u_{\lambda\bf k}^2 + v_{\lambda\bf k}^2\right)
    B_{\bf k}
\end{eqnarray} 
Using the representation $u_{\lambda\bf k}=\cosh\eta_{\lambda\bf
k}$, $v_{\lambda\bf k}=\sinh\eta_{\lambda\bf k}$, one obtains the
two branches ($\lambda =\pm1$) of the solution with
\begin{eqnarray}
    \eta_{\lambda {\bf k}}=
    \frac{1}{2}\tanh^{-1}\left(
    \frac{B_{\bf k}}{A_{\bf k}+\lambda C_{\bf k}}\right)
\end{eqnarray}
The prefactor of the remaining diagonal bilinear term in the
transformed Hamiltonian gives the spin wave energies of
Eq.~(\ref{DISPNAF}).

\bibliography{mceBSPTNS2}

\end{document}